\documentclass[twocolumn]{aastex631}

\newcommand\kms{km$\,$s$^{-1}$}
\newcommand\Msol{M$_{\odot}$}

\newcommand{\hi}{H\,{\sc i}}
\newcommand{\hii}{H\,{\sc ii}}

\begin{document}

\title{Pavo: Discovery of a star-forming dwarf galaxy just outside the Local Group\footnote{This paper includes data gathered with the 6.5~m Magellan Telescope at Las Campanas Observatory, Chile.}}

\correspondingauthor{Michael G. Jones}
\email{jonesmg@arizona.edu}

\author[0000-0002-5434-4904]{Michael G. Jones}
\affiliation{Steward Observatory, University of Arizona, 933 North Cherry Avenue, Rm. N204, Tucson, AZ 85721-0065, USA}

\author[0000-0001-9649-4815]{Bur\c{c}in Mutlu-Pakdil}
\affil{Department of Physics and Astronomy, Dartmouth College, Hanover, NH 03755, USA}

\author[0000-0003-4102-380X]{David J. Sand}
\affiliation{Steward Observatory, University of Arizona, 933 North Cherry Avenue, Rm. N204, Tucson, AZ 85721-0065, USA}

\author[0000-0001-7618-8212]{Richard Donnerstein}
\affiliation{Steward Observatory, University of Arizona, 933 North Cherry Avenue, Rm. N204, Tucson, AZ 85721-0065, USA}

\author[0000-0002-1763-4128]{Denija Crnojevi\'{c}}
\affil{University of Tampa, 401 West Kennedy Boulevard, Tampa, FL 33606, USA}

\author[0000-0001-8354-7279]{Paul Bennet}
\affiliation{Space Telescope Science Institute, 3700 San Martin Drive, Baltimore, MD 21218, USA}

\author[0000-0001-8245-779X]{Catherine E. Fielder}
\affiliation{Steward Observatory, University of Arizona, 933 North Cherry Avenue, Rm. N204, Tucson, AZ 85721-0065, USA}

\author[0000-0001-8855-3635]{Ananthan Karunakaran}
\affiliation{Department of Astronomy \& Astrophysics, University of Toronto, Toronto, ON M5S 3H4, Canada}
\affiliation{Dunlap Institute for Astronomy and Astrophysics, University of Toronto, Toronto ON, M5S 3H4, Canada}

\author[0000-0002-0956-7949]{Kristine Spekkens}
\affiliation{Department of Physics and Space Science, Royal Military College of Canada P.O. Box 17000, Station Forces Kingston, ON K7K 7B4, Canada}
\affiliation{Department of Physics, Engineering Physics and Astronomy, Queen’s University, Kingston, ON K7L 3N6, Canada}

\author[0000-0002-1468-9668]{Jay Strader}
\affil{Center for Data Intensive and Time Domain Astronomy, Department of Physics and Astronomy, Michigan State University, East Lansing, MI 48824, USA}

\author[0000-0003-1814-8620]{Ryan Urquhart}
\affil{Center for Data Intensive and Time Domain Astronomy, Department of Physics and Astronomy, Michigan State University, East Lansing, MI 48824, USA}

\author[0000-0002-5177-727X]{Dennis Zaritsky}
\affiliation{Steward Observatory, University of Arizona, 933 North Cherry Avenue, Rm. N204, Tucson, AZ 85721-0065, USA}



\begin{abstract}

We report the discovery of Pavo, a faint ($M_V = -10.0$), star-forming, irregular, and extremely isolated dwarf galaxy at $D\approx2$~Mpc. Pavo was identified in Dark Energy Camera Legacy Survey imaging via a novel approach that combines low surface brightness galaxy search algorithms and machine learning candidate classifications.
Follow-up imaging with the Inamori-Magellan Areal Camera \& Spectrograph on the 6.5~m Magellan Baade telescope revealed a color--magnitude diagram (CMD) with an old stellar population, in addition to the young population that dominates the integrated light, and a tip-of-the-red-giant-branch distance estimate of $1.99^{+0.20}_{-0.22}$~Mpc. The blue population of stars in the CMD is consistent with the youngest stars having formed no later than 150~Myr ago. We also detected no H$\alpha$ emission with SOAR telescope imaging, suggesting we may be witnessing a temporary low in Pavo's star formation. We estimate the total stellar mass of Pavo to be $\log M_\ast/\mathrm{M_\odot} = 5.6 \pm 0.2$ and measure an upper limit on its \hi \ gas mass of $1.0 \times 10^6$~\Msol \ based on the HIPASS survey. Given these properties, Pavo's closest analog is Leo~P ($D=1.6$~Mpc), previously the only known isolated, star-forming, Local Volume dwarf galaxy in this mass range. However, Pavo appears to be even more isolated, with no other known galaxy residing within over 600~kpc. As surveys and search techniques continue to improve, we anticipate an entire population of analogous objects being detected just outside the Local Group.

\end{abstract}



\section{Introduction} \label{sec:intro}

Galaxies at the extreme low-mass end ($M_\ast \lesssim 10^6$~\Msol) of the population offer unique tests of our understanding of galaxy evolution and cosmology \citep[e.g.][]{Bullock+2017,Sales+2022}. With the rise of cosmological simulations over the past few decades, several high profile discrepancies between the number and form of low-mass galaxies in these simulations and those observed in the Local Volume came to light \citep{Moore+1994,Klypin+1999,Boylan-Kolchin+2011,Pawlowski+2012,McGaugh+2012}. The inclusion of additional astrophysics in increasingly high resolution simulations \citep[e.g.][]{Brook+2015,Dutton+2016,Wetzel+2016,Sawala+2016,Garrison-Kimmel+2017} and the continued discovery of more nearby low-mass galaxies with improved surveys \citep[e.g.][]{Willman+2005,Belokurov+2006,Irwin+2007,McConnachie+2008,Koposov+2015,Drlica-Wagner+2015,Cerny+2021,Collins+2022}, have offered resolutions to most of these issues. However, our understanding of the very lowest mass galaxies, especially beyond the influence of the Milky Way (MW) and M~31, remains quite limited. In particular, current cosmological models predict that galaxies below a certain mass threshold will be permanently quenched by cosmic reionization \citep{Benson+2002,Bovill+2009,Simpson+2013,Wheeler+2015,Fitts+2017}, but this characteristic mass threshold is still largely unconstrained observationally \citep[see][for a review]{Simon+2019}.

Resolved star searches with the Sloan Digital Sky Survey \citep[SDSS,][]{York+2000}, the Dark Energy Survey \citep[DES,][]{DES}, the Dark Energy Camera Legacy Survey \citep[DECaLS,][]{Dey+2019}, and the DECam Local Volume Exploration survey \citep[DELVE,][]{Drlica-Wagner+2021} led to a flurry of ultra-faint dwarf (UFD)\footnote{Generally dwarfs fainter than $M_V = -7.7$ ($M_\ast \lesssim 10^5$~\Msol) are considered to be UFDs \citep[e.g.][]{Simon+2019}.} discoveries in (or just outside) the LG \citep[e.g.][]{Willman+2005,Belokurov+2006,Koposov+2015,Drlica-Wagner+2015,Sand+2022,Martinez-Delgado+2022,Collins+2022,Cerny+2023,McQuinn+2023b}, while focused deep imaging projects identified small samples of faint dwarfs in other nearby groups \citep[e.g.][]{Chiboucas+2013,Crnojevic+2014,Crnojevic+2016,Carlin+2016,Smercina+2017,Smercina+2018,Bennet+2019,Muller+2019,Carlin+2021,Mutlu-Pakdil+2022,McNanna+2023}. 
In the case of star-forming galaxies, many of the currently known lowest mass galaxies were identified through their \hi \ emission in the Arecibo Legacy Fast ALFA (Arecibo L-band Feed Array), or ALFALFA \citep{Giovanelli+2005,Haynes+2018}, survey and the Study of \hi \ In Extremely Low-mass Dwarfs \citep[SHIELD,][]{Cannon+2011}.


Isolated galaxies at the lowest masses are of particular interest as they sample one of the extremes of what constitutes a galaxy, while their isolation provides a clean environment in which to study their nature in the absence of influence from other more massive galaxies. At present, two of the lowest mass galaxies known in relative isolation are Leo~P, $\log L_\mathrm{V}/\mathrm{L_\ast} = 5.7$ \citep{Giovanelli+2013,McQuinn+2015b}, and Tucana~B, $\log L_\mathrm{V}/\mathrm{L_\ast} = 4.7$ \citep{Sand+2022}, with the former being star-forming and the latter quenched. Throughout this work we will use these two objects as points of comparison as they appear to bridge the threshold which determines whether or not a galaxy was quenched by cosmic reionization.


The stellar mass of Leo~P is roughly 10 times higher than that of Tucana~B (and roughly half of Leo~P's baryonic mass is in neutral gas), and in general the lowest (stellar) mass galaxies in the SHIELD sample are roughly an order of magnitude above the most massive UFDs \citep{McQuinn+2014}. Thus, there is a disconnect between the lowest mass star-forming galaxies known in the field and UFDs thought to be quenched by reionization. With suitably modified search techniques and the inclusion of machine learning it is likely that existing imaging surveys can be used to begin to fill this gap and connect these two populations. There are likely many Tucana~B and Leo~P analogs just outside the LG, where they would appear as semi-resolved objects in existing ground-based, wide-field imaging surveys. This is a regime where current search algorithms perform poorly \citep[e.g.][]{Mutlu-Pakdil+2021}. In this letter we briefly discuss a novel approach to identifying these objects and present the first notable discovery from our search.

\section{Search and discovery} \label{sec:search}

\begin{figure}
    \centering
    \includegraphics[width=\columnwidth]{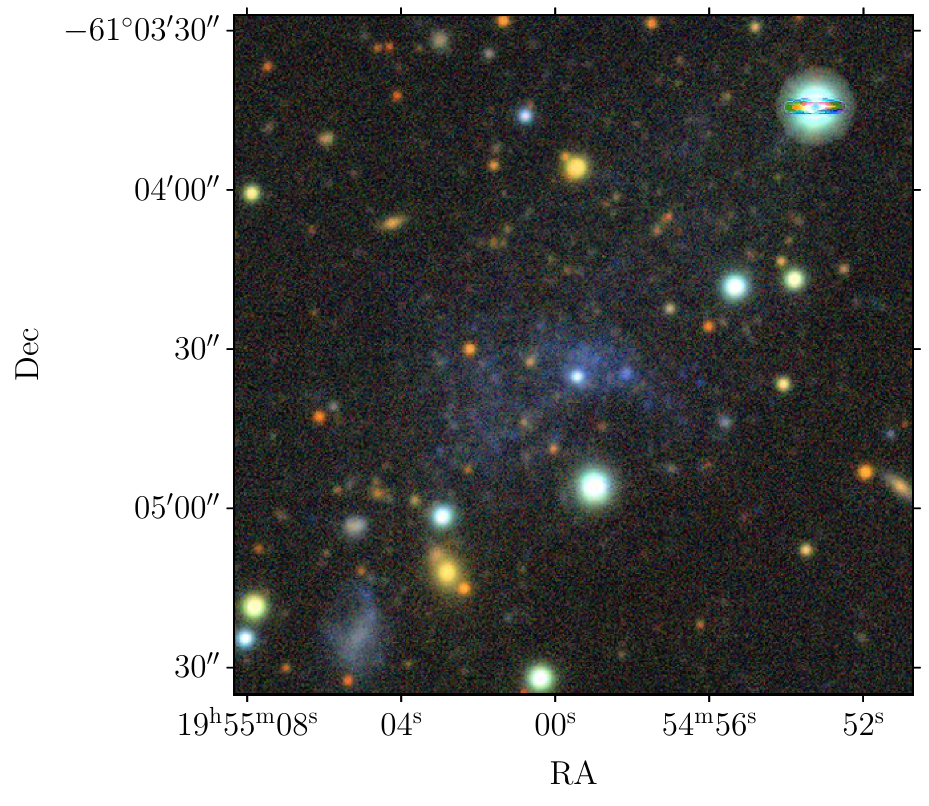}
    \caption{DECaLS DR10 $gri$ cutout of Pavo, similar to that used for the original classification.}
    \label{fig:DECaLS_cutout}
\end{figure}

Starting with publicly available images from DR10 of the Dark Energy Spectroscopic Instrument (DESI) legacy imaging surveys \citep{Dey+2019} DECaLS, the Mayall $z$-band Legacy Survey (MzLS), and the Beijing-Arizona Sky Survey (BASS), we have used a modified version of the Systematically Measuring Ultra-Diffuse Galaxies \citep[SMUDGes,][]{Zaritsky+2019} pipeline as well as retraining its neural network classifier to identify both UFD and SHIELD-like objects. Here we give a brief overview of the search process. Full details of the SMUDGes pipeline can be found in the original publications \citep{Zaritsky+2019,Zaritsky+2021,Zaritsky+2022,Zaritsky+2023} and a more complete description of our search will be presented in subsequent papers covering all candidate objects.

The original SMUDGes pipeline used the images from individual exposures, but switched to using the Legacy Survey bricks for the northern survey area. For simplicity and speed we also chose to adopt the latter strategy. The pipeline starts by identifying high surface brightness objects, subtracting them and replacing them with characteristic noise. A range of wavelet filtering scales were then used to identify candidates. Candidates must be detected in at least two bands, separated by less than 4\arcsec. The light profiles of the initial candidates were fit with \texttt{GALFIT} \citep{Peng+2002,Peng+2010}, using a forced exponential profile. Using the criteria described in \cite{Zaritsky+2023} several cuts were then applied: $r_\mathrm{e} > 4$\arcsec, $b/a > 0.34$, and $\mu_{0,g} > 22.5$~mag~arcsec$^{-2}$, or $\mu_{0,z} > 21.5$~mag~arcsec$^{-2}$ where $g$-band was unavailable. 
These cuts were originally designed to eliminate some of the large number of spurious objects picked up by the wavelet filtering, while remaining generous enough to avoid rejecting candidates that would pass a second round of fitting with full S\'{e}rsic profiles (a more computationally intensive step). However, in practice no further cuts were needed after the second round of \texttt{GALFIT} fitting (with a free S\'{e}rsic index) before beginning to classify the candidates.

These steps produced a sample of approximately $4\times10^5$ candidates over the full Legacy Survey DR10. We began by visually inspecting all candidates within a $\sim$300~deg$^{2}$ pilot area and classifying them as either candidate nearby (semi-)resolved galaxies (our target objects), other types of galaxies (not the targets of this search), or spurious detections (e.g. distant galaxy groups or Galactic cirrus). The SMUDGes convolutional neural network (CNN) image classifier was then retrained with $g$, $r$, and $z$ fits cutouts ($\sim$88\arcsec$\times$88\arcsec) of known UFDs, Local Volume dwarf galaxies, the nearest SHIELD galaxies, low luminosity globular clusters within the LG, and high confidence candidates from the pilot search. The spurious candidates were also added to the training set as negative examples. A second pilot area of $\sim$150~deg$^{2}$ near the Galactic plane was used to test the level of contamination from Galactic cirrus. This led to a second round of training containing many examples of unwanted cirrus features.

A total of 3933 candidates were given $>$90\% probability of being (semi-)resolved nearby galaxies by the CNN classifier, a factor of $\sim$100 reduction from the initial catalog of candidates.\footnote{In the case of MzLS/BASS we performed Gaussian smoothing with a 1.5 pixel kernel to reduce noise before passing the candidate cutouts to the CNN. A threshold probability of 98\% was also used for MzLS/BASS (instead of 90\%) to reduce the number of false positive candidates.} After an initial visual screening for spurious candidates that made it past the CNN (mostly cirrus, background galaxy groups, spiral arms, or tidal structures), cutouts of 1314 candidates were uploaded to the \texttt{Zooniverse} platform\footnote{\url{www.zooniverse.org}} where six volunteers from our team classified them as in the pilot search discussed above. A total of 321 of these candidates were classified by a least one volunteer as likely being a nearby, low-mass (semi-)resolved galaxy and therefore flagged for further investigation. The majority of these objects are known nearby galaxies, however, roughly a quarter are entirely uncatalogued. 

One striking example is shown in Figure~\ref{fig:DECaLS_cutout}: a blue, irregular dwarf galaxy reminiscent of Leo~P. Upon inspection, it was immediately realized that this is likely a very nearby object due to its extended and clearly semi-resolved appearance in the ground-based DECaLS imaging. We named this object ``Pavo" after the constellation in which it resides.

\section{Follow-up observations} \label{sec:follow-up}

\subsection{Magellan Baade imaging}

To further understand the physical and star formation properties of Pavo, we obtained deep $g$ and $r$-band imaging with the Inamori-Magellan Areal Camera \& Spectrograph \citep[IMACS;][]{IMACS} on the Magellan Baade 6.5-m telescope on 2023 July 14 (UT).  We used the $f/2$ camera, which has a $\sim$27.4\arcmin \ field of view and 0.2 arcsec/pixel scale.  We took 11$\times$300~s exposures in the $g$-band and 11$\times$300~s exposures in the $r$-band, with small dithers in between exposures.  The data were reduced in a standard way \citep[as in][]{Chiti20} with standard image detrending.  Astrometric correction was a two-step process, with an initial world coordinate system solution supplied by \texttt{astrometry.net} \citep{astrometry}, followed by a refined solution computed by \texttt{SCAMP} \citep{scamp}.  Final image stacking used the \texttt{SWarp} \citep{swarp} software package, using a weighted average of the input images.  The final, stacked images have point spread function full-width half maximum (FWHM) values of 0.9\arcsec \ in both the $g$ and $r$ bands.

Point source photometry was performed on the stacked IMACS images using the \texttt{daophot} and \texttt{allframe} software suite \citep{Stetson87,Stetson94}, similar to that described in \citet{Mutlu18}. We removed objects that are not point sources by culling our \texttt{allframe} catalog of outliers in $\chi$\footnote{$\chi$ is a measure of the quality of the fit between the model and the actual stars in the image.} versus magnitude, magnitude error versus magnitude, and sharpness versus magnitude space. We calibrated the photometry by matching it to the Legacy Survey DR10. We corrected for MW extinction on a star-by-star basis using the \citet{Schlegel1998} reddening maps with the coefficients from \citet{Schlafly2011}. The extinction-corrected photometry is used throughout this work.

We perform a series of artificial star tests with the \texttt{daophot} routine \texttt{addstar} to determine our photometric errors and completeness as a function of magnitude and color. Similar to \citet{Mutlu18}, we placed artificial stars into our images on a regular grid (10–20 times the image FWHM). Ten iterations are performed on the image for a total of $\sim$100,000 artificial stars each. These images are then photometered in the same way as the unaltered image stacks and the same stellar selection criteria on $\chi$, magnitude, magnitude error, and sharpness were  applied to the artificial star catalogs to determine  completeness and
magnitude uncertainties. The 50\% (90\%) completeness level is at $r$ = 26.5 (24.5) and $g$ = 27.0 (24.5) mag.

\begin{figure*}
    \centering
    \includegraphics[width=0.67\textwidth]{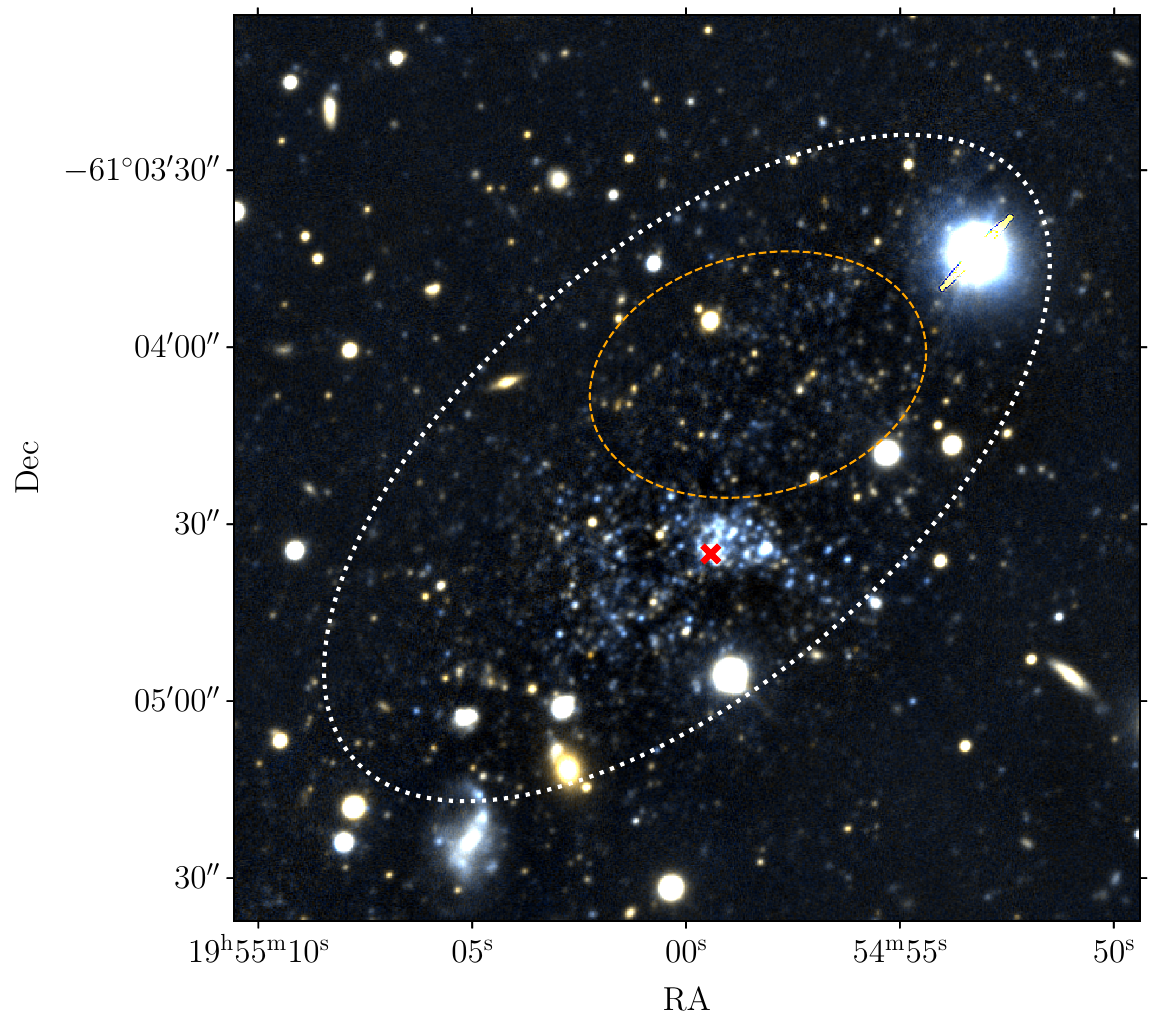}
    \caption{IMACS $g+r$ color image of Pavo. The central and SE portions of its irregular stellar body are dominated by blue stars, but it also extends to the NW where older, redder stars become the dominant population. The white dotted ellipse shows the region used to produce the full CMD of Pavo (Figure~\ref{fig:CMD}, top-left) and the orange dashed ellipse shows the region dominated by redder stars (Figure~\ref{fig:CMD}, top-right). The field contains many foreground stars, and the apparent central cluster of Pavo is actually a superimposed foreground star with detectable proper motion. We have marked this star with a red x.}
    \label{fig:RGB_image}
\end{figure*}

\subsection{SOAR H$\alpha$ imaging} \label{sec:Halpha}

To look for sites of ongoing star formation, an H$\alpha$ image of Pavo was obtained with the 4.1~m Southern Astrophysical Research (SOAR) telescope on 2023 September 17 with the Goodman Spectrograph \citep{Clemens2004}. The narrow band H$\alpha$ filter has a central wavelength of 6563~\AA \ and a width of 75~\AA. The SDSS $r$-band filter was used for the continuum image. Pavo was observed with a 3-point dither in both filters for total exposure times of 1800~s (H$\alpha$) and 180~s ($r$). Images were binned $2\times2$ to give a pixel scale of 0.3~arcsec/pixel.

Images were reduced using \texttt{ccdproc}. Individual exposures were aligned with \texttt{astroalign} before median stacking, and the final world coordinate system was derived with \texttt{astrometry.net}. Stars were extracted from both the $r$-band and H$\alpha$ images and a linear fit to the relation between their fluxes was used to scale the $r$-band image and subtract the continuum. The continuum subtraction removed the entirety of Pavo from the H$\alpha$ image, indicating that there are no sites of significant H$\alpha$ emission (i.e. \hii \ regions).

\subsection{Swift observations}

Pavo lies in a gap in the Galaxy Evolution Explorer (GALEX) all-sky imaging survey and we therefore obtained UV imaging from the NASA Neil Gehrels Swift observatory in all three UV bands (UVW1, UVM2, UVW2). Pavo was observed over two sequences on 2023 September 19 and 20. Unfortunately, the sequence taken on 2023 September 19 was unusable due to a detector artifact produced by a bright star, leaving only the 2023 September 20 sequence. Pavo was detected in all three bands, however, due to its color similarity to GALEX NUV we use only the UVM2 filter to determine star formation rate \citep{Hoversten09}. We measure the flux from Pavo using an aperture based on double the optical half light radius, masking contaminating sources such as bright stars in this region. The measured flux is very close to the zero point for the Ultra-Violet Optical Telescope (UVOT) and therefore the uncertainty on the measured flux is high. The raw flux was corrected for foreground extinction and sensitivity loss following the UVOT calibration documents. We then color correct the UVOT flux to GALEX NUV and use the relation from \cite{IglesiasParamo2006}  to derive a star formation rate (Table~\ref{tab:props}). 

\section{Properties of Pavo} \label{sec:props}

\begin{table}[]
    \centering
    \caption{Properties of Pavo}
    \begin{tabular}{lc}
    \hline\hline
    Parameter       &  Value\\ \hline
    RA              & 19:54:59.98 $\pm$ 7.9\arcsec \\
    Dec.            & $-$61:04:20.5 $\pm$ 5.8\arcsec \\
    $m_g$/mag           & $16.7\pm0.1$ \\
    $m_r$/mag           & $16.4\pm0.1$ \\
    $m_i$/mag           & $16.3\pm0.1$ \\
    $m_z$/mag           & $16.1\pm0.1$ \\
    Dist./Mpc       & $1.99^{+0.20}_{-0.22}$ \\
    $M_V$/mag           & $-10.0\pm0.1$ \\
    $r_h$           & 1.25\arcmin$\,\pm\,$0.10\arcmin \\
    $r_h$/pc        & $713 \pm 57$\\
    $\epsilon$      & $0.51\pm0.08$ \\
    $\theta$        & $131^\circ \pm 21^\circ$ \\
    $\log (\mathrm{SFR_{UV}/M_\odot \, yr^{-1}})$ & $-4.0^{+0.3}_{-0.8}$ \\
    $\log M_\ast$/\Msol  & $5.6\pm0.2$ \\
    $\log M_\mathrm{HI}$/\Msol & $<6.0$ \\
    \hline
    \end{tabular}
    \tablecomments{Stated uncertainties of distance-dependent quantities do not include the contribution from the distance uncertainty.}
    \label{tab:props}
\end{table}

\subsection{Color and morphology} \label{sec:morph}

The DECaLS image of Pavo is dominated by its young, blue stellar component, which extends from the center of Figure~\ref{fig:DECaLS_cutout} to the SE. It has a highly speckled appearance indicating that the stellar population is only slightly too distant to be resolved into individual stars. The distribution of blue stars is also highly irregular, with no clear ordered structure or center. Prior to the SOAR H$\alpha$ observations (and the non-detection of Pavo), the brightest clump near the center of the image was considered as a possible star cluster or \hii \ region. However, this object was identified as having measurable proper motion (7.2~mas/yr) in the Gaia survey \citep{Gaia,GAIADR3}, and is therefore a foreground star. \citet{Bailer-Jones+2021} estimate the distance to this star as $3.0^{+1.5}_{-1.0}$~kpc. It is marked with a red x in Figure~\ref{fig:RGB_image}.

The IMACS image of Pavo (Figure~\ref{fig:RGB_image}) resolves the stellar population, revealing that, in addition to the blue stars seen in the DECaLS image in the SE, an older, redder population extends to the NW. This is likely the underlying stellar population of the galaxy, whereas the seemingly dominant blue population only represents the most recent star formation. Star formation is highly stochastic in low-mass galaxies \citep[e.g.][]{McQuinn+2015a}, thus the blue population that dominates the total integrated light can give a very misleading impression of the underlying structure (cf. \S\ref{sec:struct}).

These findings indicate that Pavo hosts both an old stellar population and a recently formed one. It therefore cannot be a transient object, and instead is a long-lived, bona fide dwarf galaxy, albeit at a very low mass. It must also have recently contained enough gas to ignite a recent episode of star formation. It may have been continuously forming stars (perhaps with bursts and lulls) for much of its lifetime, or it may have only recently begun to form stars again after a long quiescent period. We will revisit this point and discuss the nature of its stellar population in \S\ref{sec:stellar_pop}.

\subsection{Distance} \label{sec:dist}

\begin{figure*}
    \centering
    \includegraphics[width=0.66\columnwidth]{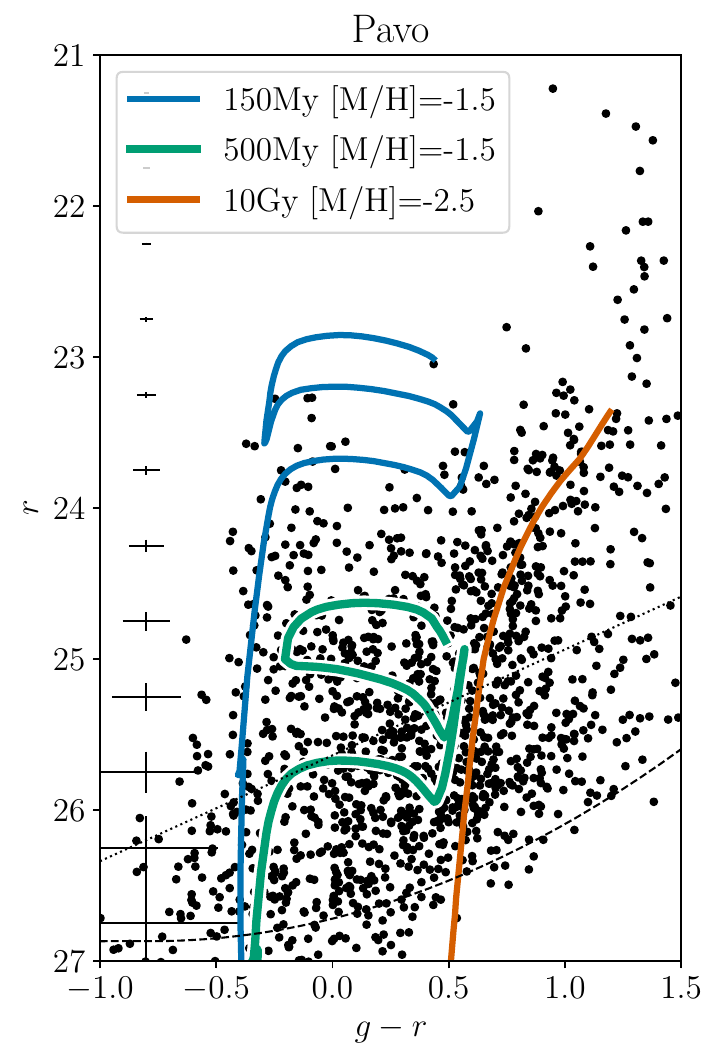}
    \includegraphics[width=0.66\columnwidth]{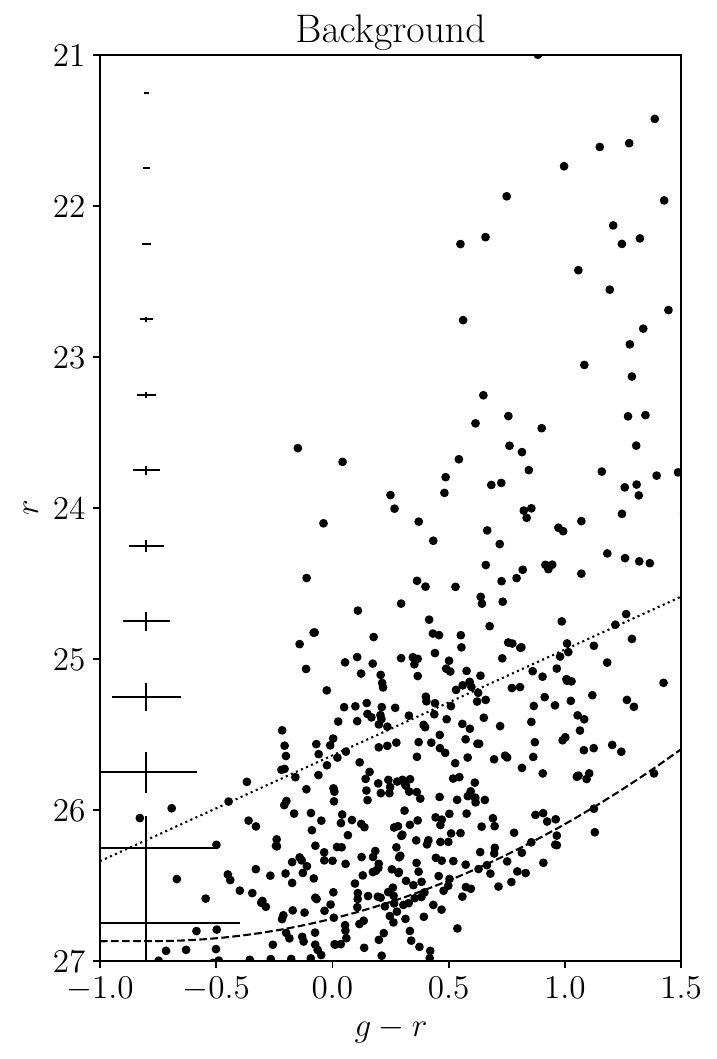}
    \includegraphics[width=0.66\columnwidth]{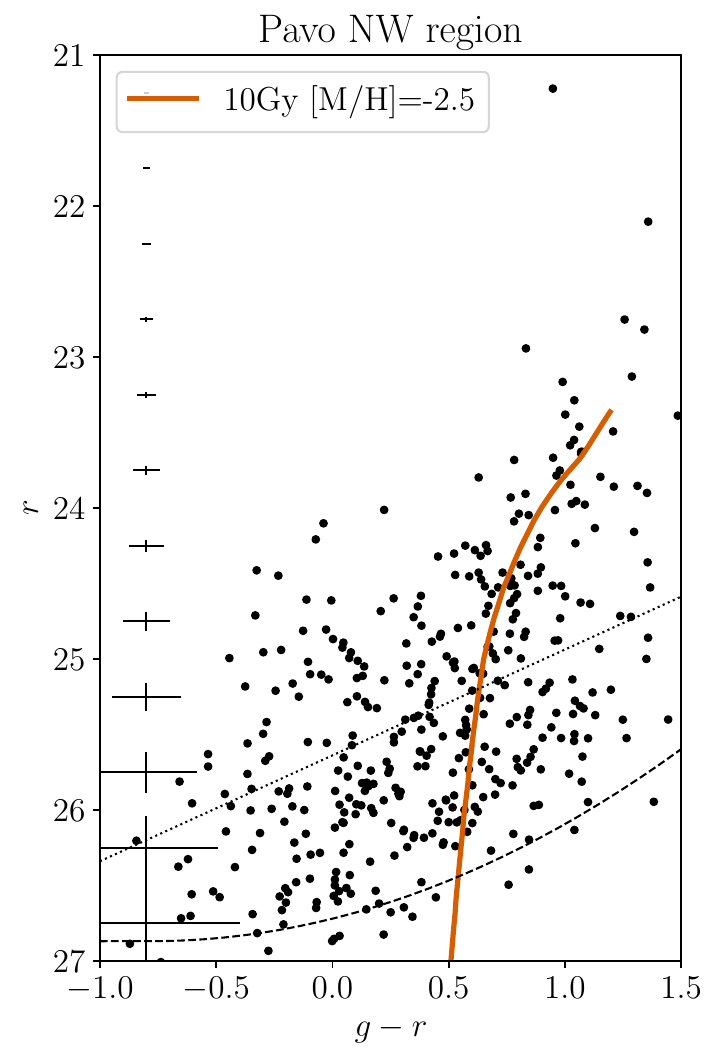}
    \includegraphics[width=0.66\columnwidth]{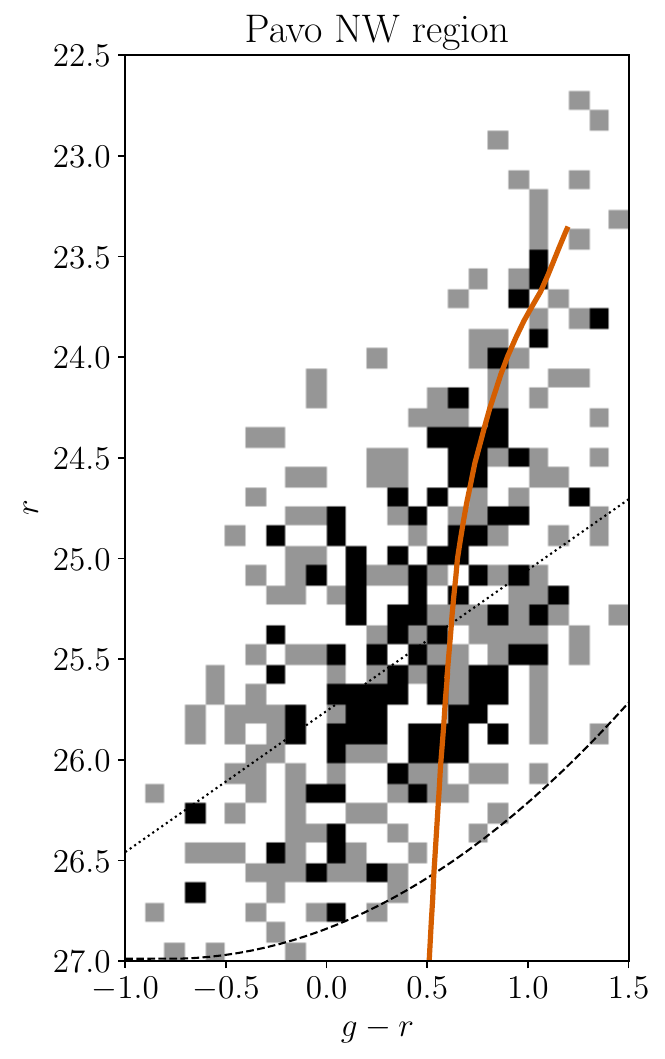}
    \includegraphics[width=0.66\columnwidth]{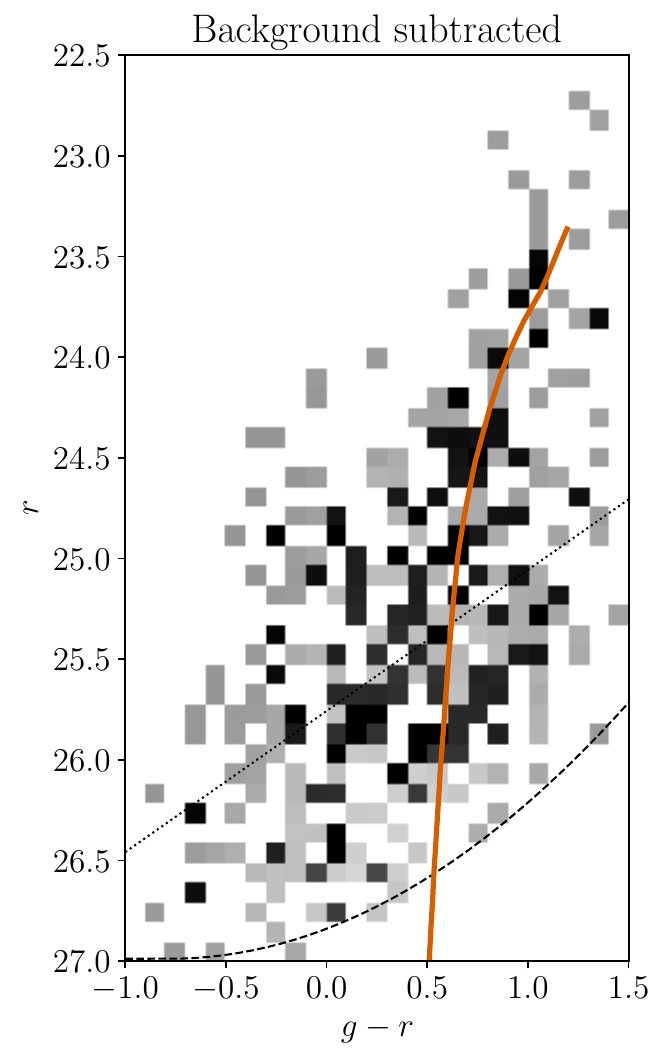}
    \includegraphics[width=0.66\columnwidth]{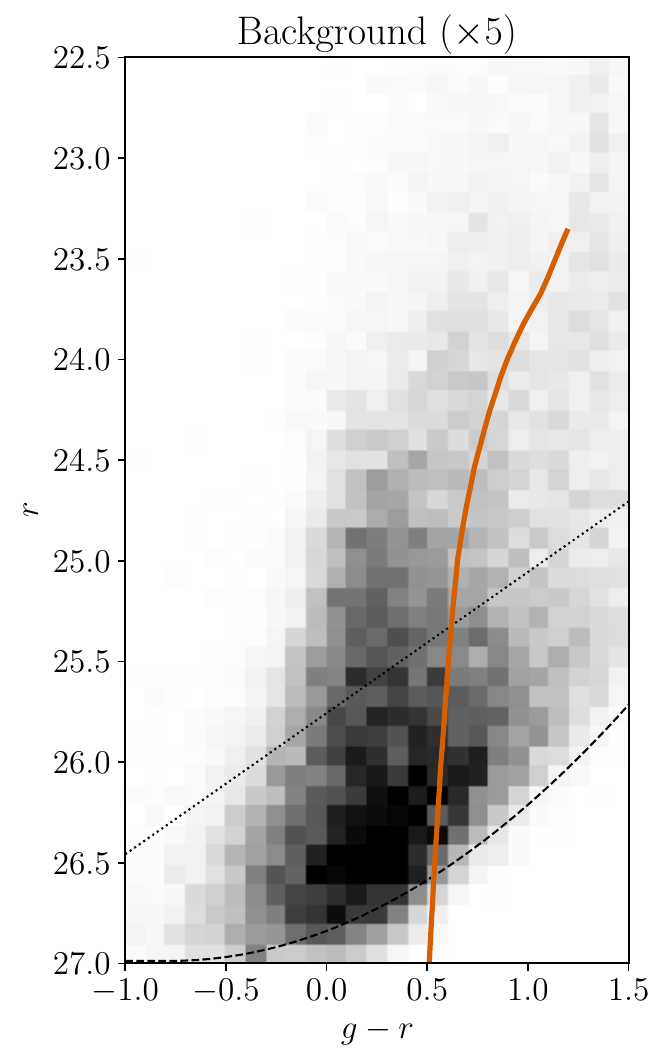}
    \caption{\textit{Top-left}: Color magnitude diagram (CMD) of the stellar population of Pavo from the IMACS imaging. The blue and green isochrones indicate the main sequence and helium burning phases of 150~Myr and 500~Myr old populations, respectively. The metallicity in both cases is set to $[M/H]=-1.5$ to roughly match that of Leo~P. The orange isochrone shows the location of an ancient and metal-poor (10~Gyr and $[M/H]=-2.5$) RGB. \textit{Top-center}: Background CMD created by downsampling the CMD of the full FoV (minus the region containing Pavo) to an area equal to the white dashed ellipse in Figure~\ref{fig:RGB_image}. \textit{Top-right}: CMD of the stellar population within the orange dotted ellipse (Figure~\ref{fig:RGB_image}). The RGB isochrone from the left panel is reproduced here. In all three panels (top) the dashed line indicates the 50\% completeness limit and the dotted line the 90\% limit, and the error bars on the left indicate the typical photometric uncertainties in bins of 0.5~mag in $r$-band. \textit{Bottom-left}: Binned version of the CMD of the NW region of Pavo (0.1~mag bins). \textit{Bottom-center}: The same CMD with the background contribution subtracted. \textit{Bottom-right}: Binned CMD of the background over the entire FoV (with Pavo excluded), weighted by five times the area of the region used in the bottom-left and bottom-center CMDs (to make it more clearly visible on the same scale). The 10~Gyr, $[M/H] = -2.5$ PARSEC isochrone (at a distance of 1.99~Mpc) is overlaid in orange.}
    \label{fig:CMD}
\end{figure*}

We derive the distance to Pavo via the tip of the red giant branch (TRGB) method \citep[e.g.,][]{dacosta90, lee93, makarov06}, where the sharp discontinuity at the bright end of an old red giant branch (RGB) population is used as a standard candle.
To perform the measurement, we only consider the NW region of the galaxy (see Figure~\ref{fig:CMD}), in order to have as little contamination as possible from the young, massive stars in Pavo; we additionally apply a color cut of $g-r<1.35$ to avoid foreground contaminants. We derive the luminosity function for the RGB stars, and fit it with a model luminosity function, after having convolved the latter with the photometric uncertainty, bias, and completeness as derived from our artificial star tests (see \citealt{crnojevic19} for details). The nonlinear least-square fitting returns a value of $r_\mathrm{TRGB}=23.48\pm0.21$ for the TRGB magnitude. This translates into a distance modulus of $(m-M)=26.49\pm0.23$~mag, or a distance of $1.99^{+0.20}_{-0.22}$~Mpc, according to the TRGB calibration in SDSS bands derived by \cite{sand14},
$M_r^\mathrm{TRGB}=-3.01\pm0.10$. An old, metal-poor RGB isochrone from the PAdova and TRieste Stellar Evolution Code \citep[PARSEC,][]{Bressan+2012}, scaled to this distance, is overlaid on a binned version of the CMD in the bottom central panel of Figure~\ref{fig:CMD}.

\subsection{Structural Parameters}\label{sec:struct}

\begin{figure}
    \centering
    \includegraphics[width=\columnwidth]{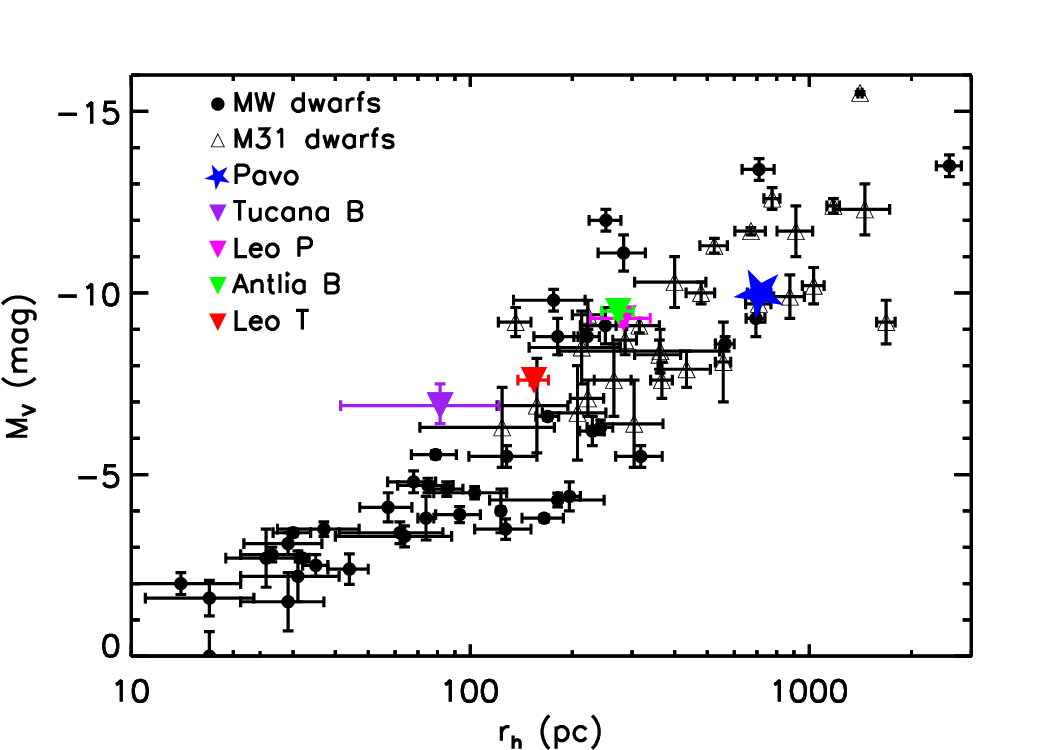}
    \caption{Pavo on the size--luminosity relation with various other Local Volume dwarfs included for comparison. Note that Pavo is at the upper limit of the scatter in effective radius (for its magnitude), while all of the highlighted comparison dwarfs are at the lower limit of the scatter.}
    \label{fig:size-lum}
\end{figure}

To measure the underlying structural parameters of Pavo, we fit an exponential profile to the two-dimensional distribution of stars consistent with the RGB using a maximum likelihood technique \citep{Martin08}, as implemented in \citet{Sand12}.  We selected RGB stars using a color--magnitude selection region within 2$\sigma$ (photometric uncertainties) of an ancient (10~Gyr) and very metal-poor ($[M/H] = -2.5$) PARSEC RGB isochrone. This isochrone is shown in orange in most panels of Figure~\ref{fig:CMD}. The top-right panel shows stars from the NW region of Pavo (orange dotted ellipse in Figure~\ref{fig:RGB_image}) compared to this isochrone. Applying this photometric selection across the whole IMACS field yielded our input RGB catalog for our structural analysis. Only stars brighter than $r$=26.5 mag were included in our selection. 

The exponential profile fit includes the central position, position angle ($\theta$), ellipticity ($\epsilon$), half-light radius ($r_\mathrm{h}$), and a constant background surface density as free parameters. The algorithm accounts for saturated foreground stars and other regions where Pavo stars could not be detected. Uncertainties were calculated through a bootstrap resampling analysis, using 1000 iterations.  As a check, we also repeated the calculations while only including RGB stars down to $r$=25.5 mag; the derived structural parameters agreed within the uncertainties.

The results of the structural analysis can be seen in Table~\ref{tab:props} and Pavo is placed on the size--luminosity relation for dwarf galaxies in Figure~\ref{fig:size-lum}. Pavo is a close match in luminosity to both Leo~P and the gas-rich dwarf Antlia~B \citep{Sand+2015b}, but is several times larger. However, all three dwarfs still fall within the scatter of the size--luminosity relation of LG dwarfs.

\subsection{Stellar population} \label{sec:stellar_pop}

\begin{figure}
    \centering
    \includegraphics[width=\columnwidth]{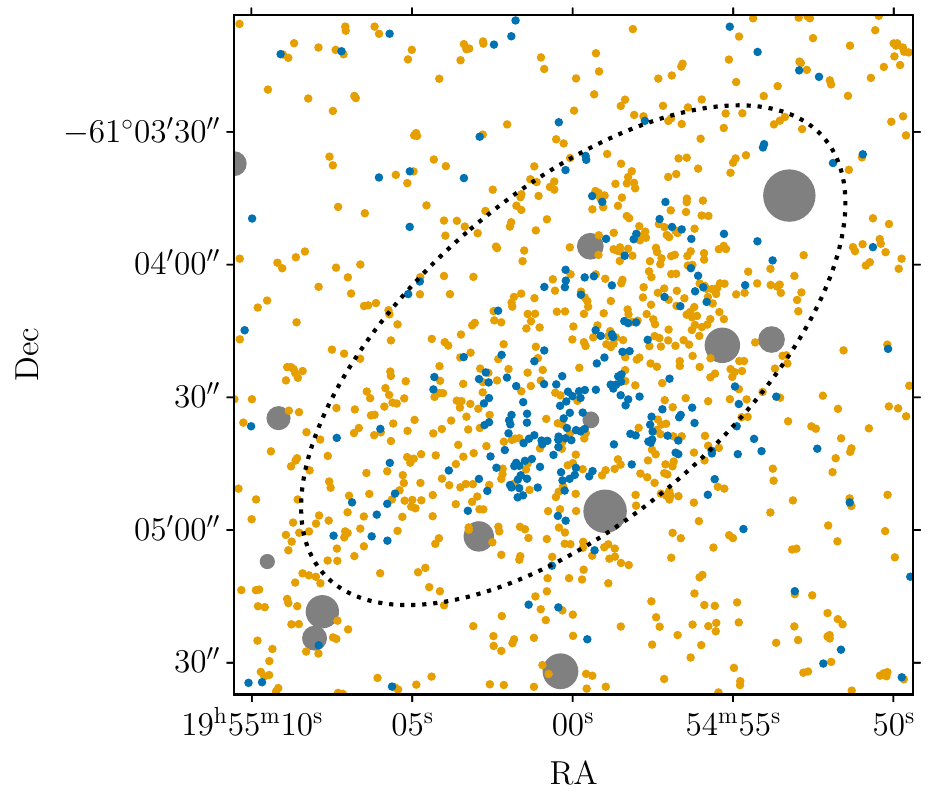}
    \caption{Map of the positions of blue, $g-r<0$ (blue points), and RGB (orange points) stars (both) brighter than $m_r = 26$. The grey points show the locations of bright stars from Gaia~DR3 ($m_G < 19.5$) with point sizes scaled by apparent brightness. The dashed black ellipse indicates the half-light radius of Pavo (Table~\ref{tab:props}). It is the same ellipse as in Figure~\ref{fig:RGB_image}.}
    \label{fig:star_map}
\end{figure}

To investigate the red and blue stellar populations in Pavo further, we selected blue stars to contrast with the RGB population from \S\ref{sec:struct}. To do this we adopted a simple selection criterion of $g-r < 0$. The positions of the red and blue populations are plotted in Figure~\ref{fig:star_map}. In both cases we plot only stars brighter than $m_r = 26$. There are holes in the distribution of stars in certain places due to bright foreground stars or background galaxies. The former are highlighted in Figure~\ref{fig:star_map} with grey circles. For the latter, we simply note that the largest gaps are in the lower-left (SE) of Figure~\ref{fig:star_map} due to bright background early-type and late-type galaxies (cf. Figure~\ref{fig:RGB_image}).

From Figure~\ref{fig:star_map} we can clearly see that the NW side of Pavo is dominated by RGB stars, while the south is dominated by blue stars. In both cases there is a clear overdensity of stars relative to the background and the majority of this overdensity is encompassed by the ellipse in Figure~\ref{fig:star_map} that was also used to produce the CMD in Figure~\ref{fig:CMD} (top-left).

As discussed in \S\ref{sec:dist}, the red stellar population in the NW of Pavo is consistent with being an ancient metal-poor RGB at $\sim$2~Mpc. This is seen even more clearly in the Hess diagram (with 0.1~mag bins) in Figure~\ref{fig:CMD} (bottom-center) where the RGB isochrone follows the densest parts of the CMD, while overlapping an underdensity in the background CMD (bottom-right).

We briefly considered the possibility that this red population might instead be merely the red side of the Helium-burning (HeB) branch, with the blue population being the blue side of the branch (and the top of the main sequence of the youngest stars). However, the fact that the two populations can be spatially separated (e.g. Figures~\ref{fig:star_map} \& \ref{fig:CMD}, top-right) refutes this notion. In addition, a relatively high metallicity (for an object of this luminosity) $[M/H] \approx -1$ would be required to match the observed colors of the two populations.

In the case of the blue population, the isochrones in Figure~\ref{fig:CMD} (top left) indicate that it is consistent with a combination of main sequence and HeB stars that are 150-500~Myr old (at 2~Mpc). The metallicity for these isochrones was chosen to approximately match that of Leo~P as we currently have no metallicity estimate for Pavo. The CMD therefore indicates that there was some star formation at least as recently as 150~Myr ago, however, when considering the brightest and bluest stars we are dealing with small number statistics and it is possible that Pavo contains some younger stars.

Our H$\alpha$ observations (\S\ref{sec:Halpha}) revealed no \hii \ regions, also supporting the notion that there are no very young stars ($<$10~Myr). However, we note that this does not mean Pavo has completely stopped forming stars. Star formation in these extremely low mass galaxies is highly stochastic and will not evenly sample the full initial mass function at any given time. It may be possible that new stars have formed in the past 10~My, but none of them are O stars and thus there are no \hii \ regions. For comparison, Leo~P has a single \hii \ region that probably contains only a single O star \citep{Rhode+2013}.

To assess this more quantitatively we consider that the Swift UV detection of Pavo indicates that over the past $\sim$200~Myr it has had an average SFR of $\sim1 \times 10^{-4}\;\mathrm{M_\ast \, yr^{-1}}$. Assuming this SFR, Pavo might have formed 1000~\Msol \ of stars in the past 10~Myr. We used PARSEC to generate 800 realizations of a 1000~\Msol, 10~Myr old single stellar population (with a Kroupa initial mass function, IMF). Every realization contained at least one O star, suggesting that the lack of H$\alpha$ emission, even at these low masses, does imply that there has been a cessation of star formation for at least the past 10~Myr. 
Another possibility is that the IMF may be top-light, which has been suggested is expected for metal-poor systems with low SFRs \citep{Pflamm-Altenburg+2009,Jerabkova+2018}. If this were the case for Pavo then it is possible that star formation is still proceeding but there are no stars massive enough to form \hii \ regions.
Overall, the simplest explanation is likely that star formation in Pavo proceeds quasi-episodically and we are currently witnessing a temporary low.

Space-based observations of Pavo will have the depth and resolution to better resolve its stellar population. This will provide considerably more information on the recent star formation, revealing if Pavo has indeed been forming stars in episodic bursts or if its SFR has been largely steady. This will give a clearer indication of whether we are witnessing a temporary lull in SFR, or the beginning of a permanent shutdown.

\subsection{Stellar and gas masses} \label{sec:mass}

To estimate the stellar mass of Pavo we performed aperture photometry (Table~\ref{tab:props}) on the DECaLS $g$, $r$, $z$, and $i$, masking clear foreground stars. The stellar mass was estimated using five different magnitude and color scaling relations (based on $r$ and $i$-band magnitudes and $g-r$ and $g-i$ colors) from \citet{Zibetti+2009}, \citet{Taylor+2011}, and \citet{Du+2020}. The median value $\log M_\ast/\mathrm{M_\odot} = 5.6$ was adopted as our stellar mass estimate and the standard deviation between the five methods, 0.2~dex, used as the uncertainty (note this does not include distance uncertainty). This mass is identical within the uncertainty to that of Leo~P, $\log M_\ast/\mathrm{M_\odot} = 5.7$ \citep{McQuinn+2015b}.

We extracted an \hi \ spectrum at the position of Pavo from the \hi \ Parkes All Sky Survey \citep[HIPASS,][]{Barnes+2001} spectral server\footnote{\url{https://www.atnf.csiro.au/research/multibeam//release/}}, but could not identify any significant emission peaks that might correspond to its neutral gas content. There is a $\sim$2$\sigma$ peak at approximately $cz_\odot = 1000$~\kms, however, unless Pavo has a particularly large peculiar velocity (e.g. $>$500~\kms) it is unlikely that this peak, if real, is associated. It is also possible that the velocity of Pavo is sufficiently small that any \hi \ emission is overwhelmed by that of the MW. Without dismissing this possibility, if we assume that Pavo's radial velocity is sufficiently large to not be blended with the \hi \ emission of the MW (e.g. $cz_\odot \gtrsim 100$~\kms, for this region of the sky), we can estimate an upper limit for its \hi \ mass based on the HIPASS spectrum. This spectrum has an rms noise of 8.2~mJy (13~\kms \ resolution). If we assume a velocity width of 30~\kms \ for Pavo \citep[Leo~P's \hi \ velocity width is 24~\kms,][]{Giovanelli+2013}, then at 2~Mpc this translates to a 3$\sigma$ upper limit on Pavo's \hi \ mass of $\log M_\mathrm{HI}/\mathrm{M_\odot} < 6.0$. We note that the \hi \ mass of Leo~P, $\log M_\ast/\mathrm{M_\odot} = 5.9$ \citep{Giovanelli+2013,McQuinn+2015b}, would be marginally below this limit.

If we go further and assume that the radial velocity of Pavo is in the range 100-500~km/s, we can also use the Galactic All-Sky Survey \citep[GASS,][]{McClure-Griffiths+2009,Kalberla+2015} to constrain Pavo's \hi \ mass. We inspected a spectral cube from GASS,\footnote{\url{https://www.astro.uni-bonn.de/hisurvey/gass/index.php}} but found no sign of \hi \ line emission from Pavo. The rms noise in this cube was 49~mK at 1~\kms \ resolution. Using 0.7~K/Jy as the approximate gain of the Parkes radio telescope, this translates to a 3$\sigma$ upper limit on the \hi \ mass of Pavo of $\log M_\mathrm{HI}/\mathrm{M_\odot} < 5.6$. Although this limit is slightly lower than that from HIPASS, the previous limit is stricter in the sense that it does not assume an upper limit of 500~\kms \ for the radial velocity of Pavo.

Although it is unclear if Pavo currently contains a significant gas reservoir it must have done recently in order to explain its population of young stars. \citet{Giovanelli+2015} suggested that very low-mass, star-forming galaxies, such as Pavo and Leo~P, might periodically become invisible to \hi \ surveys if feedback ejects much of their gas reservoirs. Recent simulations of galaxies in this mass range \citep{Rey+2020,Rey+2022} support this hypothesis. Thus, it may be that Pavo's apparent lull in SFR is also reflected in its gas content. However, deeper \hi \ observations are need to confirm/refute this possibility.

\subsection{Environment} \label{sec:env}

\begin{figure}
    \centering
    \includegraphics[width=\columnwidth]{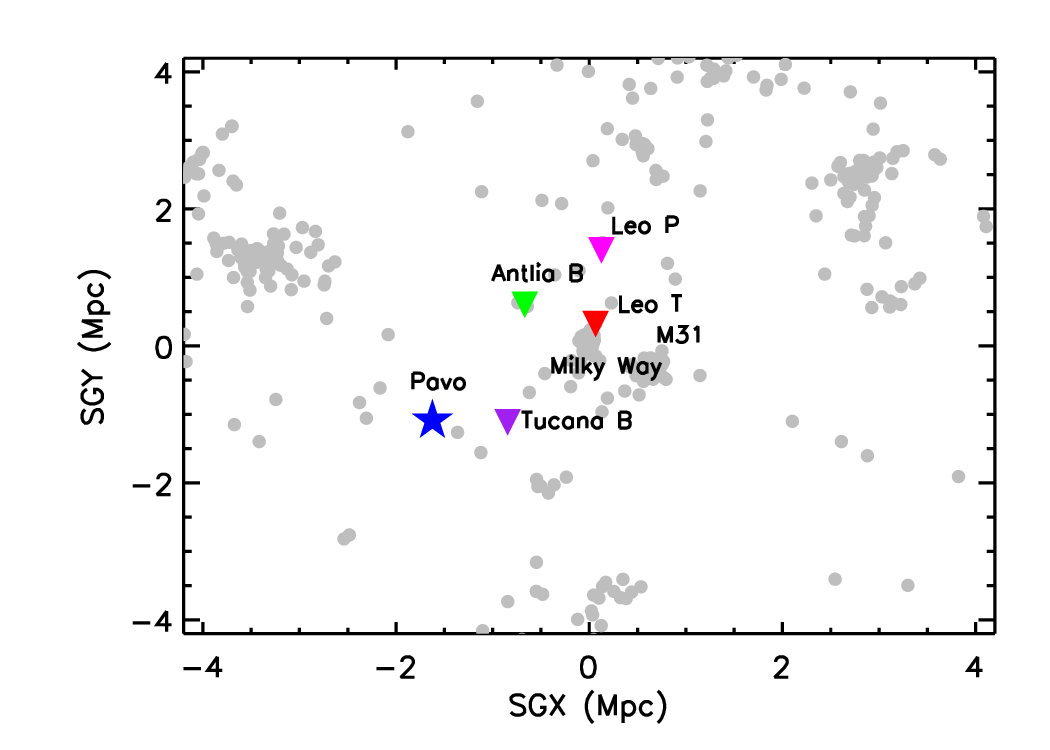}
    \includegraphics[width=\columnwidth]{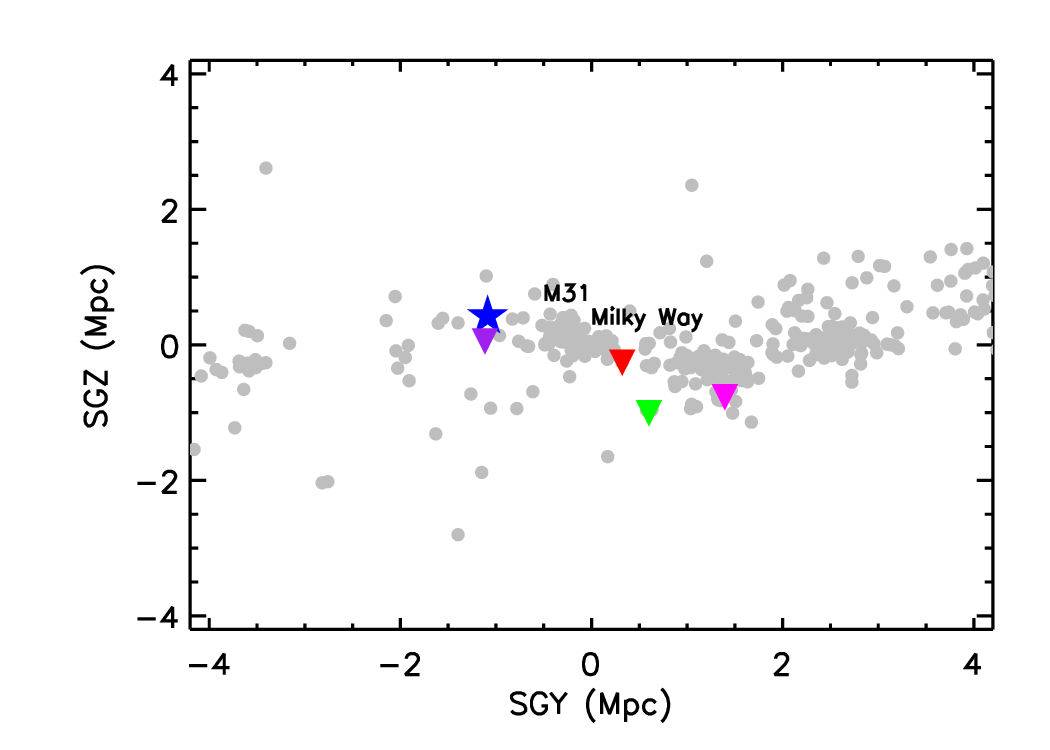}
    \caption{Supergalactic XY (top) and YZ (bottom) projections showing objects in the nearby galaxies catalog \citep{Karachentsev+2019}. Pavo is shown with a blue star. Several other notable dwarf galaxies which we refer to are highlighted for comparison. }
    \label{fig:local_env}
\end{figure}

Figure~\ref{fig:local_env} shows two projections (in Cartesian supergalactic coordinates) of nearby galaxies \citep[from][]{Karachentsev+2019}. We have highlighted Pavo as well as Tucana~B and Leo~P for the purpose of comparison. Pavo is slightly further from the MW than either Tucana~B or Leo~P, and in fact is remarkably isolated. 
The two panels of Figure~\ref{fig:local_env} show that Pavo is part of the local sheet and is in a direction away from any known nearby group/structure. 
In 3-dimensions the nearest neighbor to Pavo \citep[in the catalog of][]{Karachentsev+2019} is IC~5152 over 600~kpc away.
This makes it even more isolated than Leo~P, which is already regarded as a pristine object that likely has never entered the virial radius of a more massive system \citep{Giovanelli+2013}.

\section{Discussion} \label{sec:discussion}

The exceptional isolation of Pavo makes it an ideal candidate for follow-up observations to understand how star formation proceeds in the lowest mass star-forming galaxies. 
Upcoming space-based observations (HST-GO-17514; PI: B.~Mutlu-Pakdil) will be capable of resolving its stellar population, even in the brightest clumps where our IMACS data become too crowded (Figure~\ref{fig:star_map}), and trace both the main sequence and RGB down to fainter magnitudes. Fitting star formation histories to these deeper CMDs \citep[cf.][]{McQuinn+2015a} would indicate whether Pavo has been continuously forming stars since before reionization, formed stars episodically, or perhaps only recently revived its star formation after having been quenched at high redshift. 

Although we have compared Pavo to Leo~P and Tucana~B throughout this work, we note that really there is no other known galaxy that is a close match for all of Pavo's properties. 
Tucana~B is clearly not morphologically similar to Pavo, but we used it as a point of comparison because Tucana~B is currently the only known isolated UFD. Presumably, if Pavo were several times less massive, its gas reservoir would not have survived reionization, and today it would appear more like Tucana~B.
Pavo and Leo~P are morphologically similar and appear to have similar stellar populations, but the spatial extent of Pavo's RGB stars is roughly fives times larger than that of Leo~P's, indicating that it is quite a different object. 
The close similarity of the physical properties of Antlia~B and Leo~P (e.g. Figure~\ref{fig:size-lum}), while one is close to a massive galaxy and the other is isolated, suggests that relative isolation is unlikely to be the root cause of Pavo's more extended stellar population, and that instead there is probably an underlying, intrinsic cause.
Really what the lack of close comparisons for Pavo highlights is the dearth of isolated galaxies that are known in this mass regime. Building a better understanding of these objects and determining observationally where the tipping point between objects like Pavo and Tucana~B occurs will require a statistical sample of isolated dwarfs (both quenched and star-forming) in this mass regime ($M_\ast \lesssim 10^6$~\Msol).

Exceptionally low mass, star-forming galaxies such as Pavo and Leo~P trace the boundary of galaxy mass needed to survive cosmic reionization and continue forming stars. \citet{Tollerud+2018} used the transitional dwarf Leo~T, $\log L_\mathrm{V}/\mathrm{L_\odot} = 5.1$ and $D=0.42$~Mpc \citep{Irwin+2007}, to place an approximate upper limit on this mass scale. However, just above the mass of Leo~T there should be an abundance of Leo~P analogs just outside the LG. We posit that Pavo is merely the first such analog to be identified. By building up this sample, as well as a sample of field UFDs (e.g. Tucana~B analogs), we can constrain the mass threshold to survive reionization from both sides, without the complication of interactions with a host galaxy.

In the next few years new wide-field \hi \ surveys, currently in progress, will undoubtedly uncover new analogous star-forming dwarfs. In particular, the Commensal Radio Astronomy FasT Survey \citep[CRAFTS,][]{Zhang+2021} with the Five hundred meter Aperture Spherical Telescope (FAST) and the Widefield ASKAP (Australia Square Kilometre Array Pathfinder) L-band Legacy All-sky Blind surveY \citep[WALLABY,][]{Koribalski+2020}, will together cover almost the entire sky to at least the sensitivity of ALFALFA (several times deeper than HIPASS). These \hi \ surveys will pick out any Leo~P analogs and, at slightly higher masses, will greatly expand on the SHIELD sample. However, as these surveys will only be sensitive to gas-bearing galaxies that have fuel for star formation (and that therefore likely contain young, blue stars), many of the objects that they will uncover in this regime are likely already identifiable in Legacy Survey images, if they are searched for with a suitably optimized approach. Pavo may be the first such object to have been identified, but there are undoubtedly many more.

At optical and infra-red wavelengths the upcoming Rubin Legacy Survey of Space and Time (LSST) and the Nancy Grace Roman High Latitude Wide Area Survey will provide unprecedented opportunities to identify nearby, isolated and very low mass dwarf galaxies. Resolved star searches will be an effective tool with LSST out to a few Mpc \citep[e.g.][]{Mutlu-Pakdil+2021}, while for Roman this technique will be viable out to $\sim$10~Mpc (but over a much smaller survey area). However, as our discovery of Pavo has demonstrated, this distance range can readily be extended by incorporating machine learning-aided classification in the semi-resolved regime. As the accessible survey volume grows like the cube of distance, even increasing this maximum distance by a small factor represents a major gain in terms of the number of identifiable sources.

In subsequent papers, we will present the full sample of candidate analogs to Leo~P, SHIELD galaxies, and Tucana~B that we have uncovered in our machine learning-aided search of the DESI legacy imaging surveys. Further developing and optimizing this technique will allow semi-resolved nearby dwarf galaxies to be identified not just in existing imaging surveys, but also in upcoming next-generation surveys, such as LSST and Roman, and maximize their potential for extra-galactic science at the lowest masses.

\begin{acknowledgments}
We thank the anonymous referee for their rapid and helpful response. This work used images from the Dark Energy Camera Legacy Survey (DECaLS; Proposal ID 2014B-0404; PIs: David Schlegel and Arjun Dey). Full acknowledgment at \url{https://www.legacysurvey.org/acknowledgment/}. Based on observations obtained at the Southern Astrophysical Research (SOAR) telescope, which is a joint project of the Minist\'{e}rio da Ci\^{e}ncia, Tecnologia e Inova\c{c}\~{o}es (MCTI/LNA) do Brasil, the US National Science Foundation’s NOIRLab, the University of North Carolina at Chapel Hill (UNC), and Michigan State University (MSU).
This publication uses data generated via the \url{Zooniverse.org} platform, development of which is funded by generous support, including a Global Impact Award from Google, and by a grant from the Alfred P. Sloan Foundation.
DJS acknowledges support from NSF grants AST-1821967, 1813708 and AST-2205863.
Research by DC is supported by NSF grant AST-1814208.
KS acknowledges support from the Natural Sciences and Engineering Research Council of Canada (NSERC).
JS acknowledges support from the Packard Foundation.
AK acknowledges support from NSERC, the University of Toronto Arts \& Science Postdoctoral Fellowship program, and the Dunlap Institute.
DZ and RD acknowledge support from NSF AST-2006785 and NASA ADAP 80NSSC23K0471 for their work on the SMUDGes pipeline.
\end{acknowledgments}

%

\vspace{5mm}
\facilities{Blanco, Magellan:Baade (IMACS), SOAR, Swift, Parkes, Gaia}


\software{\href{http://astrometry.net/}{\texttt{astrometry.net}} \citep{astrometry}, \href{https://www.astromatic.net/software/scamp/}{\texttt{SCAMP}} \citep{scamp}, \href{https://www.astromatic.net/software/swarp/}{\texttt{SWarp}} \citep{swarp}, \href{https://www.astropy.org/index.html}{\texttt{astropy}} \citep{astropy2013,astropy2018}, \href{https://photutils.readthedocs.io/en/stable/}{\texttt{Photutils}} \citep{photutils}, \href{https://reproject.readthedocs.io/en/stable/}{\texttt{reproject}} \citep{reproject}, \href{https://matplotlib.org/}{\texttt{matplotlib}} \citep{matplotlib}, \href{https://numpy.org/}{\texttt{numpy}} \citep{numpy}, \href{https://scipy.org/}{\texttt{scipy}} \citep{scipy1,scipy2}, \href{https://pandas.pydata.org/}{\texttt{pandas}} \citep{pandas1,pandas2}, \href{https://astroquery.readthedocs.io/en/latest/}{\texttt{astroquery}} \citep{astroquery}, \href{https://astroalign.quatrope.org/en/latest/}{\texttt{astroalign}} \citep{astroalign}, \href{https://ccdproc.readthedocs.io/en/latest/}{ccdproc} \citep{ccdproc}, \href{https://sites.google.com/cfa.harvard.edu/saoimageds9}{\texttt{DS9}} \citep{DS9}, \href{https://users.obs.carnegiescience.edu/peng/work/galfit/galfit.html}{\texttt{GALFIT}} \citep{Peng+2002,Peng+2010}, \texttt{daophot} and \texttt{allframe} \citep{Stetson87,Stetson94}}




\bibliography{refs}{}

\begin{thebibliography}{}
\expandafter\ifx\csname natexlab\endcsname\relax\def\natexlab#1{#1}\fi
\providecommand{\url}[1]{\href{#1}{#1}}
\providecommand{\dodoi}[1]{doi:~\href{http://doi.org/#1}{\nolinkurl{#1}}}
\providecommand{\doeprint}[1]{\href{http://ascl.net/#1}{\nolinkurl{http://ascl.net/#1}}}
\providecommand{\doarXiv}[1]{\href{https://arxiv.org/abs/#1}{\nolinkurl{https://arxiv.org/abs/#1}}}

\bibitem[{{Astropy Collaboration} {et~al.}(2013){Astropy Collaboration},
  {Robitaille}, {Tollerud}, {Greenfield}, {Droettboom}, {Bray}, {Aldcroft},
  {Davis}, {Ginsburg}, {Price-Whelan}, {Kerzendorf}, {Conley}, {Crighton},
  {Barbary}, {Muna}, {Ferguson}, {Grollier}, {Parikh}, {Nair}, {Unther},
  {Deil}, {Woillez}, {Conseil}, {Kramer}, {Turner}, {Singer}, {Fox}, {Weaver},
  {Zabalza}, {Edwards}, {Azalee Bostroem}, {Burke}, {Casey}, {Crawford},
  {Dencheva}, {Ely}, {Jenness}, {Labrie}, {Lim}, {Pierfederici}, {Pontzen},
  {Ptak}, {Refsdal}, {Servillat}, \& {Streicher}}]{astropy2013}
{Astropy Collaboration}, {Robitaille}, T.~P., {Tollerud}, E.~J., {et~al.} 2013,
  \aap, 558, A33, \dodoi{10.1051/0004-6361/201322068}

\bibitem[{{Astropy Collaboration} {et~al.}(2018){Astropy Collaboration},
  {Price-Whelan}, {Sip{\H{o}}cz}, {G{\"u}nther}, {Lim}, {Crawford}, {Conseil},
  {Shupe}, {Craig}, {Dencheva}, {Ginsburg}, {VanderPlas}, {Bradley},
  {P{\'e}rez-Su{\'a}rez}, {de Val-Borro}, {Aldcroft}, {Cruz}, {Robitaille},
  {Tollerud}, {Ardelean}, {Babej}, {Bach}, {Bachetti}, {Bakanov}, {Bamford},
  {Barentsen}, {Barmby}, {Baumbach}, {Berry}, {Biscani}, {Boquien}, {Bostroem},
  {Bouma}, {Brammer}, {Bray}, {Breytenbach}, {Buddelmeijer}, {Burke},
  {Calderone}, {Cano Rodr{\'\i}guez}, {Cara}, {Cardoso}, {Cheedella}, {Copin},
  {Corrales}, {Crichton}, {D'Avella}, {Deil}, {Depagne}, {Dietrich}, {Donath},
  {Droettboom}, {Earl}, {Erben}, {Fabbro}, {Ferreira}, {Finethy}, {Fox},
  {Garrison}, {Gibbons}, {Goldstein}, {Gommers}, {Greco}, {Greenfield},
  {Groener}, {Grollier}, {Hagen}, {Hirst}, {Homeier}, {Horton}, {Hosseinzadeh},
  {Hu}, {Hunkeler}, {Ivezi{\'c}}, {Jain}, {Jenness}, {Kanarek}, {Kendrew},
  {Kern}, {Kerzendorf}, {Khvalko}, {King}, {Kirkby}, {Kulkarni}, {Kumar},
  {Lee}, {Lenz}, {Littlefair}, {Ma}, {Macleod}, {Mastropietro}, {McCully},
  {Montagnac}, {Morris}, {Mueller}, {Mumford}, {Muna}, {Murphy}, {Nelson},
  {Nguyen}, {Ninan}, {N{\"o}the}, {Ogaz}, {Oh}, {Parejko}, {Parley}, {Pascual},
  {Patil}, {Patil}, {Plunkett}, {Prochaska}, {Rastogi}, {Reddy Janga},
  {Sabater}, {Sakurikar}, {Seifert}, {Sherbert}, {Sherwood-Taylor}, {Shih},
  {Sick}, {Silbiger}, {Singanamalla}, {Singer}, {Sladen}, {Sooley},
  {Sornarajah}, {Streicher}, {Teuben}, {Thomas}, {Tremblay}, {Turner},
  {Terr{\'o}n}, {van Kerkwijk}, {de la Vega}, {Watkins}, {Weaver}, {Whitmore},
  {Woillez}, {Zabalza}, \& {Astropy Contributors}}]{astropy2018}
{Astropy Collaboration}, {Price-Whelan}, A.~M., {Sip{\H{o}}cz}, B.~M., {et~al.}
  2018, \aj, 156, 123, \dodoi{10.3847/1538-3881/aabc4f}

\bibitem[{{Bailer-Jones} {et~al.}(2021){Bailer-Jones}, {Rybizki}, {Fouesneau},
  {Demleitner}, \& {Andrae}}]{Bailer-Jones+2021}
{Bailer-Jones}, C.~A.~L., {Rybizki}, J., {Fouesneau}, M., {Demleitner}, M., \&
  {Andrae}, R. 2021, \aj, 161, 147, \dodoi{10.3847/1538-3881/abd806}

\bibitem[{{Barnes} {et~al.}(2001){Barnes}, {Staveley-Smith}, {de Blok},
  {Oosterloo}, {Stewart}, {Wright}, {Banks}, {Bhathal}, {Boyce}, {Calabretta},
  {Disney}, {Drinkwater}, {Ekers}, {Freeman}, {Gibson}, {Green}, {Haynes}, {te
  Lintel Hekkert}, {Henning}, {Jerjen}, {Juraszek}, {Kesteven}, {Kilborn},
  {Knezek}, {Koribalski}, {Kraan-Korteweg}, {Malin}, {Marquarding}, {Minchin},
  {Mould}, {Price}, {Putman}, {Ryder}, {Sadler}, {Schr{\"o}der}, {Stootman},
  {Webster}, {Wilson}, \& {Ye}}]{Barnes+2001}
{Barnes}, D.~G., {Staveley-Smith}, L., {de Blok}, W.~J.~G., {et~al.} 2001,
  \mnras, 322, 486, \dodoi{10.1046/j.1365-8711.2001.04102.x}

\bibitem[{{Belokurov} {et~al.}(2006){Belokurov}, {Zucker}, {Evans},
  {Wilkinson}, {Irwin}, {Hodgkin}, {Bramich}, {Irwin}, {Gilmore}, {Willman},
  {Vidrih}, {Newberg}, {Wyse}, {Fellhauer}, {Hewett}, {Cole}, {Bell}, {Beers},
  {Rockosi}, {Yanny}, {Grebel}, {Schneider}, {Lupton}, {Barentine},
  {Brewington}, {Brinkmann}, {Harvanek}, {Kleinman}, {Krzesinski}, {Long},
  {Nitta}, {Smith}, \& {Snedden}}]{Belokurov+2006}
{Belokurov}, V., {Zucker}, D.~B., {Evans}, N.~W., {et~al.} 2006, \apjl, 647,
  L111, \dodoi{10.1086/507324}

\bibitem[{{Bennet} {et~al.}(2019){Bennet}, {Sand}, {Crnojevi{\'c}}, {Spekkens},
  {Karunakaran}, {Zaritsky}, \& {Mutlu-Pakdil}}]{Bennet+2019}
{Bennet}, P., {Sand}, D.~J., {Crnojevi{\'c}}, D., {et~al.} 2019, \apj, 885,
  153, \dodoi{10.3847/1538-4357/ab46ab}

\bibitem[{{Benson} {et~al.}(2002){Benson}, {Frenk}, {Lacey}, {Baugh}, \&
  {Cole}}]{Benson+2002}
{Benson}, A.~J., {Frenk}, C.~S., {Lacey}, C.~G., {Baugh}, C.~M., \& {Cole}, S.
  2002, \mnras, 333, 177, \dodoi{10.1046/j.1365-8711.2002.05388.x}

\bibitem[{Beroiz {et~al.}(2020)Beroiz, Cabral, \& Sanchez}]{astroalign}
Beroiz, M., Cabral, J., \& Sanchez, B. 2020, Astronomy and Computing, 32,
  100384, \dodoi{https://doi.org/10.1016/j.ascom.2020.100384}

\bibitem[{{Bertin}(2006)}]{scamp}
{Bertin}, E. 2006, in Astronomical Society of the Pacific Conference Series,
  Vol. 351, Astronomical Data Analysis Software and Systems XV, ed.
  C.~{Gabriel}, C.~{Arviset}, D.~{Ponz}, \& S.~{Enrique}, 112

\bibitem[{{Bertin} {et~al.}(2002){Bertin}, {Mellier}, {Radovich}, {Missonnier},
  {Didelon}, \& {Morin}}]{swarp}
{Bertin}, E., {Mellier}, Y., {Radovich}, M., {et~al.} 2002, in Astronomical
  Society of the Pacific Conference Series, Vol. 281, Astronomical Data
  Analysis Software and Systems XI, ed. D.~A. {Bohlender}, D.~{Durand}, \&
  T.~H. {Handley}, 228

\bibitem[{{Bovill} \& {Ricotti}(2009)}]{Bovill+2009}
{Bovill}, M.~S., \& {Ricotti}, M. 2009, \apj, 693, 1859,
  \dodoi{10.1088/0004-637X/693/2/1859}

\bibitem[{{Boylan-Kolchin} {et~al.}(2011){Boylan-Kolchin}, {Bullock}, \&
  {Kaplinghat}}]{Boylan-Kolchin+2011}
{Boylan-Kolchin}, M., {Bullock}, J.~S., \& {Kaplinghat}, M. 2011, \mnras, 415,
  L40, \dodoi{10.1111/j.1745-3933.2011.01074.x}

\bibitem[{Bradley {et~al.}(2020)Bradley, Sip{\H o}cz, Robitaille, Tollerud,
  Vin{\'{\i}}cius, Deil, Barbary, Wilson, Busko, G{\"u}nther, Cara, Conseil,
  Bostroem, Droettboom, Bray, Bratholm, Lim, Barentsen, Craig, Pascual, Perren,
  Greco, Donath, de~Val-Borro, Kerzendorf, Bach, Weaver, D'Eugenio, Souchereau,
  \& Ferreira}]{photutils}
Bradley, L., Sip{\H o}cz, B., Robitaille, T., {et~al.} 2020, astropy/photutils:
  1.0.0, 1.0.0,  Zenodo, \dodoi{10.5281/zenodo.4044744}

\bibitem[{{Bressan} {et~al.}(2012){Bressan}, {Marigo}, {Girardi}, {Salasnich},
  {Dal Cero}, {Rubele}, \& {Nanni}}]{Bressan+2012}
{Bressan}, A., {Marigo}, P., {Girardi}, L., {et~al.} 2012, \mnras, 427, 127,
  \dodoi{10.1111/j.1365-2966.2012.21948.x}

\bibitem[{{Brook} \& {Di Cintio}(2015)}]{Brook+2015}
{Brook}, C.~B., \& {Di Cintio}, A. 2015, \mnras, 450, 3920,
  \dodoi{10.1093/mnras/stv864}

\bibitem[{{Bullock} \& {Boylan-Kolchin}(2017)}]{Bullock+2017}
{Bullock}, J.~S., \& {Boylan-Kolchin}, M. 2017, \araa, 55, 343,
  \dodoi{10.1146/annurev-astro-091916-055313}

\bibitem[{{Cannon} {et~al.}(2011){Cannon}, {Giovanelli}, {Haynes},
  {Janowiecki}, {Parker}, {Salzer}, {Adams}, {Engstrom}, {Huang}, {McQuinn},
  {Ott}, {Saintonge}, {Skillman}, {Allan}, {Erny}, {Fliss}, \&
  {Smith}}]{Cannon+2011}
{Cannon}, J.~M., {Giovanelli}, R., {Haynes}, M.~P., {et~al.} 2011, \apjl, 739,
  L22, \dodoi{10.1088/2041-8205/739/1/L22}

\bibitem[{{Carlin} {et~al.}(2016){Carlin}, {Sand}, {Price}, {Willman},
  {Karunakaran}, {Spekkens}, {Bell}, {Brodie}, {Crnojevi{\'c}}, {Forbes},
  {Hargis}, {Kirby}, {Lupton}, {Peter}, {Romanowsky}, \&
  {Strader}}]{Carlin+2016}
{Carlin}, J.~L., {Sand}, D.~J., {Price}, P., {et~al.} 2016, \apjl, 828, L5,
  \dodoi{10.3847/2041-8205/828/1/L5}

\bibitem[{{Carlin} {et~al.}(2021){Carlin}, {Mutlu-Pakdil}, {Crnojevi{\'c}},
  {Garling}, {Karunakaran}, {Peter}, {Tollerud}, {Forbes}, {Hargis}, {Lim},
  {Romanowsky}, {Sand}, {Spekkens}, \& {Strader}}]{Carlin+2021}
{Carlin}, J.~L., {Mutlu-Pakdil}, B., {Crnojevi{\'c}}, D., {et~al.} 2021, \apj,
  909, 211, \dodoi{10.3847/1538-4357/abe040}

\bibitem[{{Cerny} {et~al.}(2021){Cerny}, {Pace}, {Drlica-Wagner}, {Koposov},
  {Vivas}, {Mau}, {Riley}, {Bom}, {Carlin}, {Choi}, {Erkal}, {Ferguson},
  {James}, {Li}, {Mart{\'\i}nez-Delgado}, {Mart{\'\i}nez-V{\'a}zquez}, {Munoz},
  {Mutlu-Pakdil}, {Olsen}, {Pieres}, {Sakowska}, {Sand}, {Simon}, {Smercina},
  {Stringfellow}, {Tollerud}, {Adam{\'o}w}, {Hernandez-Lang}, {Kuropatkin},
  {Santana-Silva}, {Tucker}, {Zenteno}, \& {Delve Collaboration}}]{Cerny+2021}
{Cerny}, W., {Pace}, A.~B., {Drlica-Wagner}, A., {et~al.} 2021, \apjl, 920,
  L44, \dodoi{10.3847/2041-8213/ac2d9a}

\bibitem[{{Cerny} {et~al.}(2023){Cerny}, {Mart{\'\i}nez-V{\'a}zquez},
  {Drlica-Wagner}, {Pace}, {Mutlu-Pakdil}, {Li}, {Riley}, {Crnojevi{\'c}},
  {Bom}, {Carballo-Bello}, {Carlin}, {Chiti}, {Choi}, {Collins},
  {Darragh-Ford}, {Ferguson}, {Geha}, {Mart{\'\i}nez-Delgado}, {Massana},
  {Mau}, {Medina}, {Mu{\~n}oz}, {Nadler}, {No{\"e}l}, {Olsen}, {Pieres},
  {Sakowska}, {Simon}, {Stringfellow}, {Tollerud}, {Vivas}, {Walker},
  {Wechsler}, \& {Delve Collaboration}}]{Cerny+2023}
{Cerny}, W., {Mart{\'\i}nez-V{\'a}zquez}, C.~E., {Drlica-Wagner}, A., {et~al.}
  2023, \apj, 953, 1, \dodoi{10.3847/1538-4357/acdd78}

\bibitem[{{Chiboucas} {et~al.}(2013){Chiboucas}, {Jacobs}, {Tully}, \&
  {Karachentsev}}]{Chiboucas+2013}
{Chiboucas}, K., {Jacobs}, B.~A., {Tully}, R.~B., \& {Karachentsev}, I.~D.
  2013, \aj, 146, 126, \dodoi{10.1088/0004-6256/146/5/126}

\bibitem[{{Chiti} {et~al.}(2020){Chiti}, {Frebel}, {Jerjen}, {Kim}, \&
  {Norris}}]{Chiti20}
{Chiti}, A., {Frebel}, A., {Jerjen}, H., {Kim}, D., \& {Norris}, J.~E. 2020,
  \apj, 891, 8, \dodoi{10.3847/1538-4357/ab6d72}

\bibitem[{{Clemens} {et~al.}(2004){Clemens}, {Crain}, \&
  {Anderson}}]{Clemens2004}
{Clemens}, J.~C., {Crain}, J.~A., \& {Anderson}, R. 2004, in Society of
  Photo-Optical Instrumentation Engineers (SPIE) Conference Series, Vol. 5492,
  Ground-based Instrumentation for Astronomy, ed. A.~F.~M. {Moorwood} \&
  M.~{Iye}, 331--340, \dodoi{10.1117/12.550069}

\bibitem[{{Collins} {et~al.}(2022){Collins}, {Charles},
  {Mart{\'\i}nez-Delgado}, {Monelli}, {Karim}, {Donatiello}, {Tollerud}, \&
  {Boschin}}]{Collins+2022}
{Collins}, M. L.~M., {Charles}, E. J.~E., {Mart{\'\i}nez-Delgado}, D., {et~al.}
  2022, \mnras, 515, L72, \dodoi{10.1093/mnrasl/slac063}

\bibitem[{Craig {et~al.}(2017)Craig, Crawford, Seifert, Robitaille, Sipocz,
  Walawender, Vinícius, Ninan, Droettboom, Youn, Tollerud, Bray, walkerna22,
  Janga, stottsco, Günther, Rol, Bach, Bradley, Deil, Price-Whelan, Barbary,
  Horton, Schoenell, Nathan, Gasdia, Nelson, \& Streicher}]{ccdproc}
Craig, M., Crawford, S., Seifert, M., {et~al.} 2017, astropy/ccdproc:
  v1.3.0.post1, v1.3.0.post1,  Zenodo, \dodoi{10.5281/zenodo.1069648}

\bibitem[{{Crnojevi{\'c}} {et~al.}(2014){Crnojevi{\'c}}, {Sand}, {Caldwell},
  {Guhathakurta}, {McLeod}, {Seth}, {Simon}, {Strader}, \&
  {Toloba}}]{Crnojevic+2014}
{Crnojevi{\'c}}, D., {Sand}, D.~J., {Caldwell}, N., {et~al.} 2014, \apjl, 795,
  L35, \dodoi{10.1088/2041-8205/795/2/L35}

\bibitem[{{Crnojevi{\'c}} {et~al.}(2016){Crnojevi{\'c}}, {Sand}, {Spekkens},
  {Caldwell}, {Guhathakurta}, {McLeod}, {Seth}, {Simon}, {Strader}, \&
  {Toloba}}]{Crnojevic+2016}
{Crnojevi{\'c}}, D., {Sand}, D.~J., {Spekkens}, K., {et~al.} 2016, \apj, 823,
  19, \dodoi{10.3847/0004-637X/823/1/19}

\bibitem[{{Crnojevi{\'c}} {et~al.}(2019){Crnojevi{\'c}}, {Sand}, {Bennet},
  {Pasetto}, {Spekkens}, {Caldwell}, {Guhathakurta}, {McLeod}, {Seth}, {Simon},
  {Strader}, \& {Toloba}}]{crnojevic19}
{Crnojevi{\'c}}, D., {Sand}, D.~J., {Bennet}, P., {et~al.} 2019, \apj, 872, 80,
  \dodoi{10.3847/1538-4357/aafbe7}

\bibitem[{{Da Costa} \& {Armandroff}(1990)}]{dacosta90}
{Da Costa}, G.~S., \& {Armandroff}, T.~E. 1990, \aj, 100, 162,
  \dodoi{10.1086/115500}

\bibitem[{{Dey} {et~al.}(2019){Dey}, {Schlegel}, {Lang}, {Blum}, {Burleigh},
  {Fan}, {Findlay}, {Finkbeiner}, {Herrera}, {Juneau}, {Landriau}, {Levi},
  {McGreer}, {Meisner}, {Myers}, {Moustakas}, {Nugent}, {Patej}, {Schlafly},
  {Walker}, {Valdes}, {Weaver}, {Y{\`e}che}, {Zou}, {Zhou}, {Abareshi},
  {Abbott}, {Abolfathi}, {Aguilera}, {Alam}, {Allen}, {Alvarez}, {Annis},
  {Ansarinejad}, {Aubert}, {Beechert}, {Bell}, {BenZvi}, {Beutler}, {Bielby},
  {Bolton}, {Brice{\~n}o}, {Buckley-Geer}, {Butler}, {Calamida}, {Carlberg},
  {Carter}, {Casas}, {Castander}, {Choi}, {Comparat}, {Cukanovaite}, {Delubac},
  {DeVries}, {Dey}, {Dhungana}, {Dickinson}, {Ding}, {Donaldson}, {Duan},
  {Duckworth}, {Eftekharzadeh}, {Eisenstein}, {Etourneau}, {Fagrelius},
  {Farihi}, {Fitzpatrick}, {Font-Ribera}, {Fulmer}, {G{\"a}nsicke},
  {Gaztanaga}, {George}, {Gerdes}, {Gontcho}, {Gorgoni}, {Green}, {Guy},
  {Harmer}, {Hernandez}, {Honscheid}, {Huang}, {James}, {Jannuzi}, {Jiang},
  {Joyce}, {Karcher}, {Karkar}, {Kehoe}, {Kneib}, {Kueter-Young}, {Lan},
  {Lauer}, {Le Guillou}, {Le Van Suu}, {Lee}, {Lesser}, {Perreault Levasseur},
  {Li}, {Mann}, {Marshall}, {Mart{\'\i}nez-V{\'a}zquez}, {Martini}, {du Mas des
  Bourboux}, {McManus}, {Meier}, {M{\'e}nard}, {Metcalfe},
  {Mu{\~n}oz-Guti{\'e}rrez}, {Najita}, {Napier}, {Narayan}, {Newman}, {Nie},
  {Nord}, {Norman}, {Olsen}, {Paat}, {Palanque-Delabrouille}, {Peng},
  {Poppett}, {Poremba}, {Prakash}, {Rabinowitz}, {Raichoor}, {Rezaie},
  {Robertson}, {Roe}, {Ross}, {Ross}, {Rudnick}, {Safonova}, {Saha},
  {S{\'a}nchez}, {Savary}, {Schweiker}, {Scott}, {Seo}, {Shan}, {Silva},
  {Slepian}, {Soto}, {Sprayberry}, {Staten}, {Stillman}, {Stupak}, {Summers},
  {Sien Tie}, {Tirado}, {Vargas-Maga{\~n}a}, {Vivas}, {Wechsler}, {Williams},
  {Yang}, {Yang}, {Yapici}, {Zaritsky}, {Zenteno}, {Zhang}, {Zhang}, {Zhou}, \&
  {Zhou}}]{Dey+2019}
{Dey}, A., {Schlegel}, D.~J., {Lang}, D., {et~al.} 2019, \aj, 157, 168,
  \dodoi{10.3847/1538-3881/ab089d}

\bibitem[{{Dressler} {et~al.}(2006){Dressler}, {Hare}, {Bigelow}, \&
  {Osip}}]{IMACS}
{Dressler}, A., {Hare}, T., {Bigelow}, B.~C., \& {Osip}, D.~J. 2006, in Society
  of Photo-Optical Instrumentation Engineers (SPIE) Conference Series, Vol.
  6269, Society of Photo-Optical Instrumentation Engineers (SPIE) Conference
  Series, ed. I.~S. {McLean} \& M.~{Iye}, 62690F, \dodoi{10.1117/12.670573}

\bibitem[{{Drlica-Wagner} {et~al.}(2015){Drlica-Wagner}, {Bechtol}, {Rykoff},
  {Luque}, {Queiroz}, {Mao}, {Wechsler}, {Simon}, {Santiago}, {Yanny},
  {Balbinot}, {Dodelson}, {Fausti Neto}, {James}, {Li}, {Maia}, {Marshall},
  {Pieres}, {Stringer}, {Walker}, {Abbott}, {Abdalla}, {Allam},
  {Benoit-L{\'e}vy}, {Bernstein}, {Bertin}, {Brooks}, {Buckley-Geer}, {Burke},
  {Carnero Rosell}, {Carrasco Kind}, {Carretero}, {Crocce}, {da Costa},
  {Desai}, {Diehl}, {Dietrich}, {Doel}, {Eifler}, {Evrard}, {Finley},
  {Flaugher}, {Fosalba}, {Frieman}, {Gaztanaga}, {Gerdes}, {Gruen}, {Gruendl},
  {Gutierrez}, {Honscheid}, {Kuehn}, {Kuropatkin}, {Lahav}, {Martini},
  {Miquel}, {Nord}, {Ogando}, {Plazas}, {Reil}, {Roodman}, {Sako}, {Sanchez},
  {Scarpine}, {Schubnell}, {Sevilla-Noarbe}, {Smith}, {Soares-Santos},
  {Sobreira}, {Suchyta}, {Swanson}, {Tarle}, {Tucker}, {Vikram}, {Wester},
  {Zhang}, {Zuntz}, \& {DES Collaboration}}]{Drlica-Wagner+2015}
{Drlica-Wagner}, A., {Bechtol}, K., {Rykoff}, E.~S., {et~al.} 2015, \apj, 813,
  109, \dodoi{10.1088/0004-637X/813/2/109}

\bibitem[{{Drlica-Wagner} {et~al.}(2021){Drlica-Wagner}, {Carlin}, {Nidever},
  {Ferguson}, {Kuropatkin}, {Adam{\'o}w}, {Cerny}, {Choi}, {Esteves},
  {Mart{\'\i}nez-V{\'a}zquez}, {Mau}, {Miller}, {Mutlu-Pakdil}, {Neilsen},
  {Olsen}, {Pace}, {Riley}, {Sakowska}, {Sand}, {Santana-Silva}, {Tollerud},
  {Tucker}, {Vivas}, {Zaborowski}, {Zenteno}, {Abbott}, {Allam}, {Bechtol},
  {Bell}, {Bell}, {Bilaji}, {Bom}, {Carballo-Bello}, {Crnojevi{\'c}}, {Cioni},
  {Diaz-Ocampo}, {de Boer}, {Erkal}, {Gruendl}, {Hernandez-Lang}, {Hughes},
  {James}, {Johnson}, {Li}, {Mao}, {Mart{\'\i}nez-Delgado}, {Massana},
  {McNanna}, {Morgan}, {Nadler}, {No{\"e}l}, {Palmese}, {Peter}, {Rykoff},
  {S{\'a}nchez}, {Shipp}, {Simon}, {Smercina}, {Soares-Santos}, {Stringfellow},
  {Tavangar}, {van der Marel}, {Walker}, {Wechsler}, {Wu}, {Yanny},
  {Fitzpatrick}, {Huang}, {Jacques}, {Nikutta}, {Scott}, \& {Astro Data
  Lab}}]{Drlica-Wagner+2021}
{Drlica-Wagner}, A., {Carlin}, J.~L., {Nidever}, D.~L., {et~al.} 2021, \apjs,
  256, 2, \dodoi{10.3847/1538-4365/ac079d}

\bibitem[{{Du} \& {McGaugh}(2020)}]{Du+2020}
{Du}, W., \& {McGaugh}, S.~S. 2020, \aj, 160, 122,
  \dodoi{10.3847/1538-3881/aba754}

\bibitem[{{Dutton} {et~al.}(2016){Dutton}, {Macci{\`o}}, {Frings}, {Wang},
  {Stinson}, {Penzo}, \& {Kang}}]{Dutton+2016}
{Dutton}, A.~A., {Macci{\`o}}, A.~V., {Frings}, J., {et~al.} 2016, \mnras, 457,
  L74, \dodoi{10.1093/mnrasl/slv193}

\bibitem[{{Fitts} {et~al.}(2017){Fitts}, {Boylan-Kolchin}, {Elbert}, {Bullock},
  {Hopkins}, {O{\~n}orbe}, {Wetzel}, {Wheeler}, {Faucher-Gigu{\`e}re},
  {Kere{\v{s}}}, {Skillman}, \& {Weisz}}]{Fitts+2017}
{Fitts}, A., {Boylan-Kolchin}, M., {Elbert}, O.~D., {et~al.} 2017, \mnras, 471,
  3547, \dodoi{10.1093/mnras/stx1757}

\bibitem[{{Gaia Collaboration} {et~al.}(2016){Gaia Collaboration}, {Prusti},
  {de Bruijne}, {Brown}, {Vallenari}, {Babusiaux}, {Bailer-Jones}, {Bastian},
  {Biermann}, {Evans}, {Eyer}, {Jansen}, {Jordi}, {Klioner}, {Lammers},
  {Lindegren}, {Luri}, {Mignard}, {Milligan}, {Panem}, {Poinsignon},
  {Pourbaix}, {Randich}, {Sarri}, {Sartoretti}, {Siddiqui}, {Soubiran},
  {Valette}, {van Leeuwen}, {Walton}, {Aerts}, {Arenou}, {Cropper}, {Drimmel},
  {H{\o}g}, {Katz}, {Lattanzi}, {O'Mullane}, {Grebel}, {Holland}, {Huc},
  {Passot}, {Bramante}, {Cacciari}, {Casta{\~n}eda}, {Chaoul}, {Cheek}, {De
  Angeli}, {Fabricius}, {Guerra}, {Hern{\'a}ndez}, {Jean-Antoine-Piccolo},
  {Masana}, {Messineo}, {Mowlavi}, {Nienartowicz}, {Ord{\'o}{\~n}ez-Blanco},
  {Panuzzo}, {Portell}, {Richards}, {Riello}, {Seabroke}, {Tanga},
  {Th{\'e}venin}, {Torra}, {Els}, {Gracia-Abril}, {Comoretto},
  {Garcia-Reinaldos}, {Lock}, {Mercier}, {Altmann}, {Andrae}, {Astraatmadja},
  {Bellas-Velidis}, {Benson}, {Berthier}, {Blomme}, {Busso}, {Carry},
  {Cellino}, {Clementini}, {Cowell}, {Creevey}, {Cuypers}, {Davidson}, {De
  Ridder}, {de Torres}, {Delchambre}, {Dell'Oro}, {Ducourant}, {Fr{\'e}mat},
  {Garc{\'\i}a-Torres}, {Gosset}, {Halbwachs}, {Hambly}, {Harrison}, {Hauser},
  {Hestroffer}, {Hodgkin}, {Huckle}, {Hutton}, {Jasniewicz}, {Jordan},
  {Kontizas}, {Korn}, {Lanzafame}, {Manteiga}, {Moitinho}, {Muinonen},
  {Osinde}, {Pancino}, {Pauwels}, {Petit}, {Recio-Blanco}, {Robin}, {Sarro},
  {Siopis}, {Smith}, {Smith}, {Sozzetti}, {Thuillot}, {van Reeven}, {Viala},
  {Abbas}, {Abreu Aramburu}, {Accart}, {Aguado}, {Allan}, {Allasia},
  {Altavilla}, {{\'A}lvarez}, {Alves}, {Anderson}, {Andrei}, {Anglada Varela},
  {Antiche}, {Antoja}, {Ant{\'o}n}, {Arcay}, {Atzei}, {Ayache}, {Bach},
  {Baker}, {Balaguer-N{\'u}{\~n}ez}, {Barache}, {Barata}, {Barbier}, {Barblan},
  {Baroni}, {Barrado y Navascu{\'e}s}, {Barros}, {Barstow}, {Becciani},
  {Bellazzini}, {Bellei}, {Bello Garc{\'\i}a}, {Belokurov}, {Bendjoya},
  {Berihuete}, {Bianchi}, {Bienaym{\'e}}, {Billebaud}, {Blagorodnova},
  {Blanco-Cuaresma}, {Boch}, {Bombrun}, {Borrachero}, {Bouquillon}, {Bourda},
  {Bouy}, {Bragaglia}, {Breddels}, {Brouillet}, {Br{\"u}semeister},
  {Bucciarelli}, {Budnik}, {Burgess}, {Burgon}, {Burlacu}, {Busonero}, {Buzzi},
  {Caffau}, {Cambras}, {Campbell}, {Cancelliere}, {Cantat-Gaudin}, {Carlucci},
  {Carrasco}, {Castellani}, {Charlot}, {Charnas}, {Charvet}, {Chassat},
  {Chiavassa}, {Clotet}, {Cocozza}, {Collins}, {Collins}, {Costigan}, {Crifo},
  {Cross}, {Crosta}, {Crowley}, {Dafonte}, {Damerdji}, {Dapergolas}, {David},
  {David}, {De Cat}, {de Felice}, {de Laverny}, {De Luise}, {De March}, {de
  Martino}, {de Souza}, {Debosscher}, {del Pozo}, {Delbo}, {Delgado},
  {Delgado}, {di Marco}, {Di Matteo}, {Diakite}, {Distefano}, {Dolding}, {Dos
  Anjos}, {Drazinos}, {Dur{\'a}n}, {Dzigan}, {Ecale}, {Edvardsson}, {Enke},
  {Erdmann}, {Escolar}, {Espina}, {Evans}, {Eynard Bontemps}, {Fabre},
  {Fabrizio}, {Faigler}, {Falc{\~a}o}, {Farr{\`a}s Casas}, {Faye}, {Federici},
  {Fedorets}, {Fern{\'a}ndez-Hern{\'a}ndez}, {Fernique}, {Fienga}, {Figueras},
  {Filippi}, {Findeisen}, {Fonti}, {Fouesneau}, {Fraile}, {Fraser}, {Fuchs},
  {Furnell}, {Gai}, {Galleti}, {Galluccio}, {Garabato}, {Garc{\'\i}a-Sedano},
  {Gar{\'e}}, {Garofalo}, {Garralda}, {Gavras}, {Gerssen}, {Geyer}, {Gilmore},
  {Girona}, {Giuffrida}, {Gomes}, {Gonz{\'a}lez-Marcos},
  {Gonz{\'a}lez-N{\'u}{\~n}ez}, {Gonz{\'a}lez-Vidal}, {Granvik}, {Guerrier},
  {Guillout}, {Guiraud}, {G{\'u}rpide}, {Guti{\'e}rrez-S{\'a}nchez}, {Guy},
  {Haigron}, {Hatzidimitriou}, {Haywood}, {Heiter}, {Helmi}, {Hobbs},
  {Hofmann}, {Holl}, {Holland}, {Hunt}, {Hypki}, {Icardi}, {Irwin}, {Jevardat
  de Fombelle}, {Jofr{\'e}}, {Jonker}, {Jorissen}, {Julbe}, {Karampelas},
  {Kochoska}, {Kohley}, {Kolenberg}, {Kontizas}, {Koposov}, {Kordopatis},
  {Koubsky}, {Kowalczyk}, {Krone-Martins}, {Kudryashova}, {Kull}, {Bachchan},
  {Lacoste-Seris}, {Lanza}, {Lavigne}, {Le Poncin-Lafitte}, {Lebreton},
  {Lebzelter}, {Leccia}, {Leclerc}, {Lecoeur-Taibi}, {Lemaitre}, {Lenhardt},
  {Leroux}, {Liao}, {Licata}, {Lindstr{\o}m}, {Lister}, {Livanou}, {Lobel},
  {L{\"o}ffler}, {L{\'o}pez}, {Lopez-Lozano}, {Lorenz}, {Loureiro},
  {MacDonald}, {Magalh{\~a}es Fernandes}, {Managau}, {Mann}, {Mantelet},
  {Marchal}, {Marchant}, {Marconi}, {Marie}, {Marinoni}, {Marrese},
  {Marschalk{\'o}}, {Marshall}, {Mart{\'\i}n-Fleitas}, {Martino}, {Mary},
  {Matijevi{\v{c}}}, {Mazeh}, {McMillan}, {Messina}, {Mestre}, {Michalik},
  {Millar}, {Miranda}, {Molina}, {Molinaro}, {Molinaro}, {Moln{\'a}r},
  {Moniez}, {Montegriffo}, {Monteiro}, {Mor}, {Mora}, {Morbidelli}, {Morel},
  {Morgenthaler}, {Morley}, {Morris}, {Mulone}, {Muraveva}, {Musella},
  {Narbonne}, {Nelemans}, {Nicastro}, {Noval}, {Ord{\'e}novic},
  {Ordieres-Mer{\'e}}, {Osborne}, {Pagani}, {Pagano}, {Pailler}, {Palacin},
  {Palaversa}, {Parsons}, {Paulsen}, {Pecoraro}, {Pedrosa}, {Pentik{\"a}inen},
  {Pereira}, {Pichon}, {Piersimoni}, {Pineau}, {Plachy}, {Plum}, {Poujoulet},
  {Pr{\v{s}}a}, {Pulone}, {Ragaini}, {Rago}, {Rambaux}, {Ramos-Lerate},
  {Ranalli}, {Rauw}, {Read}, {Regibo}, {Renk}, {Reyl{\'e}}, {Ribeiro},
  {Rimoldini}, {Ripepi}, {Riva}, {Rixon}, {Roelens}, {Romero-G{\'o}mez},
  {Rowell}, {Royer}, {Rudolph}, {Ruiz-Dern}, {Sadowski}, {Sagrist{\`a}
  Sell{\'e}s}, {Sahlmann}, {Salgado}, {Salguero}, {Sarasso}, {Savietto},
  {Schnorhk}, {Schultheis}, {Sciacca}, {Segol}, {Segovia}, {Segransan},
  {Serpell}, {Shih}, {Smareglia}, {Smart}, {Smith}, {Solano}, {Solitro},
  {Sordo}, {Soria Nieto}, {Souchay}, {Spagna}, {Spoto}, {Stampa}, {Steele},
  {Steidelm{\"u}ller}, {Stephenson}, {Stoev}, {Suess}, {S{\"u}veges}, {Surdej},
  {Szabados}, {Szegedi-Elek}, {Tapiador}, {Taris}, {Tauran}, {Taylor},
  {Teixeira}, {Terrett}, {Tingley}, {Trager}, {Turon}, {Ulla}, {Utrilla},
  {Valentini}, {van Elteren}, {Van Hemelryck}, {van Leeuwen}, {Varadi},
  {Vecchiato}, {Veljanoski}, {Via}, {Vicente}, {Vogt}, {Voss}, {Votruba},
  {Voutsinas}, {Walmsley}, {Weiler}, {Weingrill}, {Werner}, {Wevers},
  {Whitehead}, {Wyrzykowski}, {Yoldas}, {{\v{Z}}erjal}, {Zucker}, {Zurbach},
  {Zwitter}, {Alecu}, {Allen}, {Allende Prieto}, {Amorim},
  {Anglada-Escud{\'e}}, {Arsenijevic}, {Azaz}, {Balm}, {Beck}, {Bernstein},
  {Bigot}, {Bijaoui}, {Blasco}, {Bonfigli}, {Bono}, {Boudreault}, {Bressan},
  {Brown}, {Brunet}, {Bunclark}, {Buonanno}, {Butkevich}, {Carret}, {Carrion},
  {Chemin}, {Ch{\'e}reau}, {Corcione}, {Darmigny}, {de Boer}, {de Teodoro}, {de
  Zeeuw}, {Delle Luche}, {Domingues}, {Dubath}, {Fodor}, {Fr{\'e}zouls},
  {Fries}, {Fustes}, {Fyfe}, {Gallardo}, {Gallegos}, {Gardiol}, {Gebran},
  {Gomboc}, {G{\'o}mez}, {Grux}, {Gueguen}, {Heyrovsky}, {Hoar}, {Iannicola},
  {Isasi Parache}, {Janotto}, {Joliet}, {Jonckheere}, {Keil}, {Kim},
  {Klagyivik}, {Klar}, {Knude}, {Kochukhov}, {Kolka}, {Kos}, {Kutka}, {Lainey},
  {LeBouquin}, {Liu}, {Loreggia}, {Makarov}, {Marseille}, {Martayan},
  {Martinez-Rubi}, {Massart}, {Meynadier}, {Mignot}, {Munari}, {Nguyen},
  {Nordlander}, {Ocvirk}, {O'Flaherty}, {Olias Sanz}, {Ortiz}, {Osorio},
  {Oszkiewicz}, {Ouzounis}, {Palmer}, {Park}, {Pasquato}, {Peltzer}, {Peralta},
  {P{\'e}turaud}, {Pieniluoma}, {Pigozzi}, {Poels}, {Prat}, {Prod'homme},
  {Raison}, {Rebordao}, {Risquez}, {Rocca-Volmerange}, {Rosen}, {Ruiz-Fuertes},
  {Russo}, {Sembay}, {Serraller Vizcaino}, {Short}, {Siebert}, {Silva},
  {Sinachopoulos}, {Slezak}, {Soffel}, {Sosnowska}, {Strai{\v{z}}ys}, {ter
  Linden}, {Terrell}, {Theil}, {Tiede}, {Troisi}, {Tsalmantza}, {Tur},
  {Vaccari}, {Vachier}, {Valles}, {Van Hamme}, {Veltz}, {Virtanen}, {Wallut},
  {Wichmann}, {Wilkinson}, {Ziaeepour}, \& {Zschocke}}]{Gaia}
{Gaia Collaboration}, {Prusti}, T., {de Bruijne}, J.~H.~J., {et~al.} 2016,
  \aap, 595, A1, \dodoi{10.1051/0004-6361/201629272}

\bibitem[{{Gaia Collaboration} {et~al.}(2023){Gaia Collaboration}, {Vallenari},
  {Brown}, {Prusti}, {de Bruijne}, {Arenou}, {Babusiaux}, {Biermann},
  {Creevey}, {Ducourant}, {Evans}, {Eyer}, {Guerra}, {Hutton}, {Jordi},
  {Klioner}, {Lammers}, {Lindegren}, {Luri}, {Mignard}, {Panem}, {Pourbaix},
  {Randich}, {Sartoretti}, {Soubiran}, {Tanga}, {Walton}, {Bailer-Jones},
  {Bastian}, {Drimmel}, {Jansen}, {Katz}, {Lattanzi}, {van Leeuwen}, {Bakker},
  {Cacciari}, {Casta{\~n}eda}, {De Angeli}, {Fabricius}, {Fouesneau},
  {Fr{\'e}mat}, {Galluccio}, {Guerrier}, {Heiter}, {Masana}, {Messineo},
  {Mowlavi}, {Nicolas}, {Nienartowicz}, {Pailler}, {Panuzzo}, {Riclet}, {Roux},
  {Seabroke}, {Sordo}, {Th{\'e}venin}, {Gracia-Abril}, {Portell}, {Teyssier},
  {Altmann}, {Andrae}, {Audard}, {Bellas-Velidis}, {Benson}, {Berthier},
  {Blomme}, {Burgess}, {Busonero}, {Busso}, {C{\'a}novas}, {Carry}, {Cellino},
  {Cheek}, {Clementini}, {Damerdji}, {Davidson}, {de Teodoro}, {Nu{\~n}ez
  Campos}, {Delchambre}, {Dell'Oro}, {Esquej}, {Fern{\'a}ndez-Hern{\'a}ndez},
  {Fraile}, {Garabato}, {Garc{\'\i}a-Lario}, {Gosset}, {Haigron}, {Halbwachs},
  {Hambly}, {Harrison}, {Hern{\'a}ndez}, {Hestroffer}, {Hodgkin}, {Holl},
  {Jan{\ss}en}, {Jevardat de Fombelle}, {Jordan}, {Krone-Martins}, {Lanzafame},
  {L{\"o}ffler}, {Marchal}, {Marrese}, {Moitinho}, {Muinonen}, {Osborne},
  {Pancino}, {Pauwels}, {Recio-Blanco}, {Reyl{\'e}}, {Riello}, {Rimoldini},
  {Roegiers}, {Rybizki}, {Sarro}, {Siopis}, {Smith}, {Sozzetti}, {Utrilla},
  {van Leeuwen}, {Abbas}, {{\'A}brah{\'a}m}, {Abreu Aramburu}, {Aerts},
  {Aguado}, {Ajaj}, {Aldea-Montero}, {Altavilla}, {{\'A}lvarez}, {Alves},
  {Anders}, {Anderson}, {Anglada Varela}, {Antoja}, {Baines}, {Baker},
  {Balaguer-N{\'u}{\~n}ez}, {Balbinot}, {Balog}, {Barache}, {Barbato},
  {Barros}, {Barstow}, {Bartolom{\'e}}, {Bassilana}, {Bauchet}, {Becciani},
  {Bellazzini}, {Berihuete}, {Bernet}, {Bertone}, {Bianchi}, {Binnenfeld},
  {Blanco-Cuaresma}, {Blazere}, {Boch}, {Bombrun}, {Bossini}, {Bouquillon},
  {Bragaglia}, {Bramante}, {Breedt}, {Bressan}, {Brouillet}, {Brugaletta},
  {Bucciarelli}, {Burlacu}, {Butkevich}, {Buzzi}, {Caffau}, {Cancelliere},
  {Cantat-Gaudin}, {Carballo}, {Carlucci}, {Carnerero}, {Carrasco},
  {Casamiquela}, {Castellani}, {Castro-Ginard}, {Chaoul}, {Charlot}, {Chemin},
  {Chiaramida}, {Chiavassa}, {Chornay}, {Comoretto}, {Contursi}, {Cooper},
  {Cornez}, {Cowell}, {Crifo}, {Cropper}, {Crosta}, {Crowley}, {Dafonte},
  {Dapergolas}, {David}, {David}, {de Laverny}, {De Luise}, {De March}, {De
  Ridder}, {de Souza}, {de Torres}, {del Peloso}, {del Pozo}, {Delbo},
  {Delgado}, {Delisle}, {Demouchy}, {Dharmawardena}, {Di Matteo}, {Diakite},
  {Diener}, {Distefano}, {Dolding}, {Edvardsson}, {Enke}, {Fabre}, {Fabrizio},
  {Faigler}, {Fedorets}, {Fernique}, {Fienga}, {Figueras}, {Fournier},
  {Fouron}, {Fragkoudi}, {Gai}, {Garcia-Gutierrez}, {Garcia-Reinaldos},
  {Garc{\'\i}a-Torres}, {Garofalo}, {Gavel}, {Gavras}, {Gerlach}, {Geyer},
  {Giacobbe}, {Gilmore}, {Girona}, {Giuffrida}, {Gomel}, {Gomez},
  {Gonz{\'a}lez-N{\'u}{\~n}ez}, {Gonz{\'a}lez-Santamar{\'\i}a},
  {Gonz{\'a}lez-Vidal}, {Granvik}, {Guillout}, {Guiraud},
  {Guti{\'e}rrez-S{\'a}nchez}, {Guy}, {Hatzidimitriou}, {Hauser}, {Haywood},
  {Helmer}, {Helmi}, {Sarmiento}, {Hidalgo}, {Hilger}, {H{\l}adczuk}, {Hobbs},
  {Holland}, {Huckle}, {Jardine}, {Jasniewicz}, {Jean-Antoine Piccolo},
  {Jim{\'e}nez-Arranz}, {Jorissen}, {Juaristi Campillo}, {Julbe}, {Karbevska},
  {Kervella}, {Khanna}, {Kontizas}, {Kordopatis}, {Korn}, {K{\'o}sp{\'a}l},
  {Kostrzewa-Rutkowska}, {Kruszy{\'n}ska}, {Kun}, {Laizeau}, {Lambert},
  {Lanza}, {Lasne}, {Le Campion}, {Lebreton}, {Lebzelter}, {Leccia}, {Leclerc},
  {Lecoeur-Taibi}, {Liao}, {Licata}, {Lindstr{\o}m}, {Lister}, {Livanou},
  {Lobel}, {Lorca}, {Loup}, {Madrero Pardo}, {Magdaleno Romeo}, {Managau},
  {Mann}, {Manteiga}, {Marchant}, {Marconi}, {Marcos}, {Marcos Santos},
  {Mar{\'\i}n Pina}, {Marinoni}, {Marocco}, {Marshall}, {Martin Polo},
  {Mart{\'\i}n-Fleitas}, {Marton}, {Mary}, {Masip}, {Massari},
  {Mastrobuono-Battisti}, {Mazeh}, {McMillan}, {Messina}, {Michalik}, {Millar},
  {Mints}, {Molina}, {Molinaro}, {Moln{\'a}r}, {Monari}, {Mongui{\'o}},
  {Montegriffo}, {Montero}, {Mor}, {Mora}, {Morbidelli}, {Morel}, {Morris},
  {Muraveva}, {Murphy}, {Musella}, {Nagy}, {Noval}, {Oca{\~n}a}, {Ogden},
  {Ordenovic}, {Osinde}, {Pagani}, {Pagano}, {Palaversa}, {Palicio},
  {Pallas-Quintela}, {Panahi}, {Payne-Wardenaar}, {Pe{\~n}alosa Esteller},
  {Penttil{\"a}}, {Pichon}, {Piersimoni}, {Pineau}, {Plachy}, {Plum}, {Poggio},
  {Pr{\v{s}}a}, {Pulone}, {Racero}, {Ragaini}, {Rainer}, {Raiteri}, {Rambaux},
  {Ramos}, {Ramos-Lerate}, {Re Fiorentin}, {Regibo}, {Richards}, {Rios Diaz},
  {Ripepi}, {Riva}, {Rix}, {Rixon}, {Robichon}, {Robin}, {Robin}, {Roelens},
  {Rogues}, {Rohrbasser}, {Romero-G{\'o}mez}, {Rowell}, {Royer}, {Ruz Mieres},
  {Rybicki}, {Sadowski}, {S{\'a}ez N{\'u}{\~n}ez}, {Sagrist{\`a} Sell{\'e}s},
  {Sahlmann}, {Salguero}, {Samaras}, {Sanchez Gimenez}, {Sanna},
  {Santove{\~n}a}, {Sarasso}, {Schultheis}, {Sciacca}, {Segol}, {Segovia},
  {S{\'e}gransan}, {Semeux}, {Shahaf}, {Siddiqui}, {Siebert}, {Siltala},
  {Silvelo}, {Slezak}, {Slezak}, {Smart}, {Snaith}, {Solano}, {Solitro},
  {Souami}, {Souchay}, {Spagna}, {Spina}, {Spoto}, {Steele},
  {Steidelm{\"u}ller}, {Stephenson}, {S{\"u}veges}, {Surdej}, {Szabados},
  {Szegedi-Elek}, {Taris}, {Taylor}, {Teixeira}, {Tolomei}, {Tonello}, {Torra},
  {Torra}, {Torralba Elipe}, {Trabucchi}, {Tsounis}, {Turon}, {Ulla}, {Unger},
  {Vaillant}, {van Dillen}, {van Reeven}, {Vanel}, {Vecchiato}, {Viala},
  {Vicente}, {Voutsinas}, {Weiler}, {Wevers}, {Wyrzykowski}, {Yoldas}, {Yvard},
  {Zhao}, {Zorec}, {Zucker}, \& {Zwitter}}]{GAIADR3}
{Gaia Collaboration}, {Vallenari}, A., {Brown}, A.~G.~A., {et~al.} 2023, \aap,
  674, A1, \dodoi{10.1051/0004-6361/202243940}

\bibitem[{{Garrison-Kimmel} {et~al.}(2017){Garrison-Kimmel}, {Wetzel},
  {Bullock}, {Hopkins}, {Boylan-Kolchin}, {Faucher-Gigu{\`e}re}, {Kere{\v{s}}},
  {Quataert}, {Sanderson}, {Graus}, \& {Kelley}}]{Garrison-Kimmel+2017}
{Garrison-Kimmel}, S., {Wetzel}, A., {Bullock}, J.~S., {et~al.} 2017, \mnras,
  471, 1709, \dodoi{10.1093/mnras/stx1710}

\bibitem[{{Ginsburg} {et~al.}(2019){Ginsburg}, {Sip{\H{o}}cz}, {Brasseur},
  {Cowperthwaite}, {Craig}, {Deil}, {Guillochon}, {Guzman}, {Liedtke}, {Lian
  Lim}, {Lockhart}, {Mommert}, {Morris}, {Norman}, {Parikh}, {Persson},
  {Robitaille}, {Segovia}, {Singer}, {Tollerud}, {de Val-Borro}, {Valtchanov},
  {Woillez}, {Astroquery Collaboration}, \& {a subset of astropy
  Collaboration}}]{astroquery}
{Ginsburg}, A., {Sip{\H{o}}cz}, B.~M., {Brasseur}, C.~E., {et~al.} 2019, \aj,
  157, 98, \dodoi{10.3847/1538-3881/aafc33}

\bibitem[{{Giovanelli} \& {Haynes}(2015)}]{Giovanelli+2015}
{Giovanelli}, R., \& {Haynes}, M.~P. 2015, \aapr, 24, 1,
  \dodoi{10.1007/s00159-015-0085-3}

\bibitem[{{Giovanelli} {et~al.}(2005){Giovanelli}, {Haynes}, {Kent},
  {Perillat}, {Saintonge}, {Brosch}, {Catinella}, {Hoffman}, {Stierwalt},
  {Spekkens}, {Lerner}, {Masters}, {Momjian}, {Rosenberg}, {Springob},
  {Boselli}, {Charmandaris}, {Darling}, {Davies}, {Garcia Lambas}, {Gavazzi},
  {Giovanardi}, {Hardy}, {Hunt}, {Iovino}, {Karachentsev}, {Karachentseva},
  {Koopmann}, {Marinoni}, {Minchin}, {Muller}, {Putman}, {Pantoja}, {Salzer},
  {Scodeggio}, {Skillman}, {Solanes}, {Valotto}, {van Driel}, \& {van
  Zee}}]{Giovanelli+2005}
{Giovanelli}, R., {Haynes}, M.~P., {Kent}, B.~R., {et~al.} 2005, \aj, 130,
  2598, \dodoi{10.1086/497431}

\bibitem[{{Giovanelli} {et~al.}(2013){Giovanelli}, {Haynes}, {Adams}, {Cannon},
  {Rhode}, {Salzer}, {Skillman}, {Bernstein-Cooper}, \&
  {McQuinn}}]{Giovanelli+2013}
{Giovanelli}, R., {Haynes}, M.~P., {Adams}, E. A.~K., {et~al.} 2013, \aj, 146,
  15, \dodoi{10.1088/0004-6256/146/1/15}

\bibitem[{{Haynes} {et~al.}(2018){Haynes}, {Giovanelli}, {Kent}, {Adams},
  {Balonek}, {Craig}, {Fertig}, {Finn}, {Giovanardi}, {Hallenbeck}, {Hess},
  {Hoffman}, {Huang}, {Jones}, {Koopmann}, {Kornreich}, {Leisman}, {Miller},
  {Moorman}, {O'Connor}, {O'Donoghue}, {Papastergis}, {Troischt}, {Stark}, \&
  {Xiao}}]{Haynes+2018}
{Haynes}, M.~P., {Giovanelli}, R., {Kent}, B.~R., {et~al.} 2018, \apj, 861, 49,
  \dodoi{10.3847/1538-4357/aac956}

\bibitem[{{Hoversten} {et~al.}(2009){Hoversten}, {Gronwall}, {Vanden Berk},
  {Koch}, {Breeveld}, {Curran}, {Hinshaw}, {Marshall}, {Roming}, {Siegel}, \&
  {Still}}]{Hoversten09}
{Hoversten}, E.~A., {Gronwall}, C., {Vanden Berk}, D.~E., {et~al.} 2009, \apj,
  705, 1462, \dodoi{10.1088/0004-637X/705/2/1462}

\bibitem[{{Hunter}(2007)}]{matplotlib}
{Hunter}, J.~D. 2007, Computing in Science and Engineering, 9, 90,
  \dodoi{10.1109/MCSE.2007.55}

\bibitem[{{Iglesias-P{\'a}ramo} {et~al.}(2006){Iglesias-P{\'a}ramo}, {Buat},
  {Takeuchi}, {Xu}, {Boissier}, {Boselli}, {Burgarella}, {Madore}, {Gil de
  Paz}, {Bianchi}, {Barlow}, {Byun}, {Donas}, {Forster}, {Friedman}, {Heckman},
  {Jelinski}, {Lee}, {Malina}, {Martin}, {Milliard}, {Morrissey}, {Neff},
  {Rich}, {Schiminovich}, {Seibert}, {Siegmund}, {Small}, {Szalay}, {Welsh}, \&
  {Wyder}}]{IglesiasParamo2006}
{Iglesias-P{\'a}ramo}, J., {Buat}, V., {Takeuchi}, T.~T., {et~al.} 2006, \apjs,
  164, 38, \dodoi{10.1086/502628}

\bibitem[{{Irwin} {et~al.}(2007){Irwin}, {Belokurov}, {Evans}, {Ryan-Weber},
  {de Jong}, {Koposov}, {Zucker}, {Hodgkin}, {Gilmore}, {Prema}, {Hebb},
  {Begum}, {Fellhauer}, {Hewett}, {Kennicutt}, {Wilkinson}, {Bramich},
  {Vidrih}, {Rix}, {Beers}, {Barentine}, {Brewington}, {Harvanek},
  {Krzesinski}, {Long}, {Nitta}, \& {Snedden}}]{Irwin+2007}
{Irwin}, M.~J., {Belokurov}, V., {Evans}, N.~W., {et~al.} 2007, \apjl, 656,
  L13, \dodoi{10.1086/512183}

\bibitem[{{Je{\v{r}}{\'a}bkov{\'a}} {et~al.}(2018){Je{\v{r}}{\'a}bkov{\'a}},
  {Hasani Zonoozi}, {Kroupa}, {Beccari}, {Yan}, {Vazdekis}, \&
  {Zhang}}]{Jerabkova+2018}
{Je{\v{r}}{\'a}bkov{\'a}}, T., {Hasani Zonoozi}, A., {Kroupa}, P., {et~al.}
  2018, \aap, 620, A39, \dodoi{10.1051/0004-6361/201833055}

\bibitem[{{Joye} \& {Mandel}(2003)}]{DS9}
{Joye}, W.~A., \& {Mandel}, E. 2003, in Astronomical Society of the Pacific
  Conference Series, Vol. 295, Astronomical Data Analysis Software and Systems
  XII, ed. H.~E. {Payne}, R.~I. {Jedrzejewski}, \& R.~N. {Hook}, 489

\bibitem[{{Kalberla} \& {Haud}(2015)}]{Kalberla+2015}
{Kalberla}, P.~M.~W., \& {Haud}, U. 2015, \aap, 578, A78,
  \dodoi{10.1051/0004-6361/201525859}

\bibitem[{{Karachentsev} \& {Kaisina}(2019)}]{Karachentsev+2019}
{Karachentsev}, I.~D., \& {Kaisina}, E.~I. 2019, Astrophysical Bulletin, 74,
  111, \dodoi{10.1134/S1990341319020019}

\bibitem[{{Klypin} {et~al.}(1999){Klypin}, {Kravtsov}, {Valenzuela}, \&
  {Prada}}]{Klypin+1999}
{Klypin}, A., {Kravtsov}, A.~V., {Valenzuela}, O., \& {Prada}, F. 1999, \apj,
  522, 82, \dodoi{10.1086/307643}

\bibitem[{{Koposov} {et~al.}(2015){Koposov}, {Belokurov}, {Torrealba}, \&
  {Evans}}]{Koposov+2015}
{Koposov}, S.~E., {Belokurov}, V., {Torrealba}, G., \& {Evans}, N.~W. 2015,
  \apj, 805, 130, \dodoi{10.1088/0004-637X/805/2/130}

\bibitem[{{Koribalski} {et~al.}(2020){Koribalski}, {Staveley-Smith},
  {Westmeier}, {Serra}, {Spekkens}, {Wong}, {Lee-Waddell}, {Lagos},
  {Obreschkow}, {Ryan-Weber}, {Zwaan}, {Kilborn}, {Bekiaris}, {Bekki},
  {Bigiel}, {Boselli}, {Bosma}, {Catinella}, {Chauhan}, {Cluver}, {Colless},
  {Courtois}, {Crain}, {de Blok}, {D{\'e}nes}, {Duffy}, {Elagali}, {Fluke},
  {For}, {Heald}, {Henning}, {Hess}, {Holwerda}, {Howlett}, {Jarrett}, {Jones},
  {Jones}, {J{\'o}zsa}, {Jurek}, {J{\"u}tte}, {Kamphuis}, {Karachentsev},
  {Kerp}, {Kleiner}, {Kraan-Korteweg}, {L{\'o}pez-S{\'a}nchez}, {Madrid},
  {Meyer}, {Mould}, {Murugeshan}, {Norris}, {Oh}, {Oosterloo}, {Popping},
  {Putman}, {Reynolds}, {Rhee}, {Robotham}, {Ryder}, {Schr{\"o}der}, {Shao},
  {Stevens}, {Taylor}, {van{\^A} der Hulst}, {Verdes-Montenegro}, {Wakker},
  {Wang}, {Whiting}, {Winkel}, \& {Wolf}}]{Koribalski+2020}
{Koribalski}, B.~S., {Staveley-Smith}, L., {Westmeier}, T., {et~al.} 2020,
  \apss, 365, 118, \dodoi{10.1007/s10509-020-03831-4}

\bibitem[{{Lang} {et~al.}(2010){Lang}, {Hogg}, {Mierle}, {Blanton}, \&
  {Roweis}}]{astrometry}
{Lang}, D., {Hogg}, D.~W., {Mierle}, K., {Blanton}, M., \& {Roweis}, S. 2010,
  \aj, 139, 1782, \dodoi{10.1088/0004-6256/139/5/1782}

\bibitem[{{Lee} {et~al.}(1993){Lee}, {Freedman}, \& {Madore}}]{lee93}
{Lee}, M.~G., {Freedman}, W.~L., \& {Madore}, B.~F. 1993, \apj, 417, 553,
  \dodoi{10.1086/173334}

\bibitem[{{Makarov} {et~al.}(2006){Makarov}, {Makarova}, {Rizzi}, {Tully},
  {Dolphin}, {Sakai}, \& {Shaya}}]{makarov06}
{Makarov}, D., {Makarova}, L., {Rizzi}, L., {et~al.} 2006, \aj, 132, 2729,
  \dodoi{10.1086/508925}

\bibitem[{{Martin} {et~al.}(2008){Martin}, {de Jong}, \& {Rix}}]{Martin08}
{Martin}, N.~F., {de Jong}, J. T.~A., \& {Rix}, H.-W. 2008, \apj, 684, 1075,
  \dodoi{10.1086/590336}

\bibitem[{{Mart{\'\i}nez-Delgado} {et~al.}(2022){Mart{\'\i}nez-Delgado},
  {Karim}, {Charles}, {Boschin}, {Monelli}, {Collins}, {Donatiello}, \&
  {Alfaro}}]{Martinez-Delgado+2022}
{Mart{\'\i}nez-Delgado}, D., {Karim}, N., {Charles}, E. J.~E., {et~al.} 2022,
  \mnras, 509, 16, \dodoi{10.1093/mnras/stab2797}

\bibitem[{{McClure-Griffiths} {et~al.}(2009){McClure-Griffiths}, {Pisano},
  {Calabretta}, {Ford}, {Lockman}, {Staveley-Smith}, {Kalberla}, {Bailin},
  {Dedes}, {Janowiecki}, {Gibson}, {Murphy}, {Nakanishi}, \&
  {Newton-McGee}}]{McClure-Griffiths+2009}
{McClure-Griffiths}, N.~M., {Pisano}, D.~J., {Calabretta}, M.~R., {et~al.}
  2009, \apjs, 181, 398, \dodoi{10.1088/0067-0049/181/2/398}

\bibitem[{{McConnachie} {et~al.}(2008){McConnachie}, {Huxor}, {Martin},
  {Irwin}, {Chapman}, {Fahlman}, {Ferguson}, {Ibata}, {Lewis}, {Richer}, \&
  {Tanvir}}]{McConnachie+2008}
{McConnachie}, A.~W., {Huxor}, A., {Martin}, N.~F., {et~al.} 2008, \apj, 688,
  1009, \dodoi{10.1086/591313}

\bibitem[{{McGaugh}(2012)}]{McGaugh+2012}
{McGaugh}, S.~S. 2012, \aj, 143, 40, \dodoi{10.1088/0004-6256/143/2/40}

\bibitem[{{McNanna} {et~al.}(2023){McNanna}, {Bechtol}, {Mau}, {Nadler},
  {Medoff}, {Drlica-Wagner}, {Cerny}, {Crnojevic}, {Mutlu-Pakdil}, {Vivas},
  {Pace}, {Carlin}, {Collins}, {Martinez-Delgado}, {Martinez-Vazquez}, {Noel},
  {Riley}, {Sand}, {Smercina}, {Wechsler}, {Abbott}, {Aguena}, {Alves},
  {Bacon}, {Bom}, {Brooks}, {Burke}, {Carballo-Bello}, {Carnero Rosell},
  {Carretero}, {da Costa}, {Davis}, {De Vicente}, {Diehl}, {Doel}, {Ferrero},
  {Frieman}, {Giannini}, {Gruen}, {Gutierrez}, {Gruendl}, {Hinton},
  {Hollowood}, {Honscheid}, {James}, {Kuehn}, {Marshall}, {Mena-Fernandez},
  {Miquel}, {Pereira}, {Pieres}, {Plazas Malagon}, {Sakowska}, {Sanchez},
  {Sanchez Cid}, {Santiago}, {Sevilla-Noarbe}, {Smith}, {Stringfellow},
  {Suchyta}, {Swanson}, {Tarle}, {Weaverdyck}, \& {Wiseman}}]{McNanna+2023}
{McNanna}, M., {Bechtol}, K., {Mau}, S., {et~al.} 2023, arXiv e-prints,
  arXiv:2309.04467, \dodoi{10.48550/arXiv.2309.04467}

\bibitem[{{McQuinn} {et~al.}(2023){McQuinn}, {Mao}, {Cohen}, {Shih}, {Buckley},
  \& {Dolphin}}]{McQuinn+2023b}
{McQuinn}, K. B.~W., {Mao}, Y.-Y., {Cohen}, R.~E., {et~al.} 2023, arXiv
  e-prints, arXiv:2307.08738, \dodoi{10.48550/arXiv.2307.08738}

\bibitem[{{McQuinn} {et~al.}(2014){McQuinn}, {Cannon}, {Dolphin}, {Skillman},
  {Salzer}, {Haynes}, {Adams}, {Cave}, {Elson}, {Giovanelli}, {Ott}, \&
  {Saintonge}}]{McQuinn+2014}
{McQuinn}, K. B.~W., {Cannon}, J.~M., {Dolphin}, A.~E., {et~al.} 2014, \apj,
  785, 3, \dodoi{10.1088/0004-637X/785/1/3}

\bibitem[{{McQuinn} {et~al.}(2015{\natexlab{a}}){McQuinn}, {Skillman},
  {Dolphin}, {Cannon}, {Salzer}, {Rhode}, {Adams}, {Berg}, {Giovanelli},
  {Girardi}, \& {Haynes}}]{McQuinn+2015b}
{McQuinn}, K. B.~W., {Skillman}, E.~D., {Dolphin}, A., {et~al.}
  2015{\natexlab{a}}, \apj, 812, 158, \dodoi{10.1088/0004-637X/812/2/158}

\bibitem[{{McQuinn} {et~al.}(2015{\natexlab{b}}){McQuinn}, {Cannon}, {Dolphin},
  {Skillman}, {Haynes}, {Simones}, {Salzer}, {Adams}, {Elson}, {Giovanelli}, \&
  {Ott}}]{McQuinn+2015a}
{McQuinn}, K. B.~W., {Cannon}, J.~M., {Dolphin}, A.~E., {et~al.}
  2015{\natexlab{b}}, \apj, 802, 66, \dodoi{10.1088/0004-637X/802/1/66}

\bibitem[{{Millman} \& {Aivazis}(2011)}]{scipy2}
{Millman}, K.~J., \& {Aivazis}, M. 2011, Computing in Science and Engineering,
  13, 9, \dodoi{10.1109/MCSE.2011.36}

\bibitem[{{Moore}(1994)}]{Moore+1994}
{Moore}, B. 1994, \nat, 370, 629, \dodoi{10.1038/370629a0}

\bibitem[{{M{\"u}ller} {et~al.}(2019){M{\"u}ller}, {Rejkuba}, {Pawlowski},
  {Ibata}, {Lelli}, {Hilker}, \& {Jerjen}}]{Muller+2019}
{M{\"u}ller}, O., {Rejkuba}, M., {Pawlowski}, M.~S., {et~al.} 2019, \aap, 629,
  A18, \dodoi{10.1051/0004-6361/201935807}

\bibitem[{{Mutlu-Pakdil} {et~al.}(2018){Mutlu-Pakdil}, {Sand}, {Carlin},
  {Spekkens}, {Caldwell}, {Crnojevi{\'c}}, {Hughes}, {Willman}, \&
  {Zaritsky}}]{Mutlu18}
{Mutlu-Pakdil}, B., {Sand}, D.~J., {Carlin}, J.~L., {et~al.} 2018, \apj, 863,
  25, \dodoi{10.3847/1538-4357/aacd0e}

\bibitem[{{Mutlu-Pakdil} {et~al.}(2021){Mutlu-Pakdil}, {Sand}, {Crnojevi{\'c}},
  {Drlica-Wagner}, {Caldwell}, {Guhathakurta}, {Seth}, {Simon}, {Strader}, \&
  {Toloba}}]{Mutlu-Pakdil+2021}
{Mutlu-Pakdil}, B., {Sand}, D.~J., {Crnojevi{\'c}}, D., {et~al.} 2021, \apj,
  918, 88, \dodoi{10.3847/1538-4357/ac0db8}

\bibitem[{{Mutlu-Pakdil} {et~al.}(2022){Mutlu-Pakdil}, {Sand}, {Crnojevi{\'c}},
  {Jones}, {Caldwell}, {Guhathakurta}, {Seth}, {Simon}, {Spekkens}, {Strader},
  \& {Toloba}}]{Mutlu-Pakdil+2022}
---. 2022, \apj, 926, 77, \dodoi{10.3847/1538-4357/ac4418}

\bibitem[{{Oliphant}(2007)}]{scipy1}
{Oliphant}, T.~E. 2007, Computing in Science and Engineering, 9, 10,
  \dodoi{10.1109/MCSE.2007.58}

\bibitem[{pandas~development team(2020)}]{pandas2}
pandas~development team, T. 2020, pandas-dev/pandas: Pandas, latest,  Zenodo,
  \dodoi{10.5281/zenodo.3509134}

\bibitem[{{Pawlowski} {et~al.}(2012){Pawlowski}, {Pflamm-Altenburg}, \&
  {Kroupa}}]{Pawlowski+2012}
{Pawlowski}, M.~S., {Pflamm-Altenburg}, J., \& {Kroupa}, P. 2012, \mnras, 423,
  1109, \dodoi{10.1111/j.1365-2966.2012.20937.x}

\bibitem[{{Peng} {et~al.}(2002){Peng}, {Ho}, {Impey}, \& {Rix}}]{Peng+2002}
{Peng}, C.~Y., {Ho}, L.~C., {Impey}, C.~D., \& {Rix}, H.-W. 2002, \aj, 124,
  266, \dodoi{10.1086/340952}

\bibitem[{{Peng} {et~al.}(2010){Peng}, {Ho}, {Impey}, \& {Rix}}]{Peng+2010}
---. 2010, \aj, 139, 2097, \dodoi{10.1088/0004-6256/139/6/2097}

\bibitem[{{Pflamm-Altenburg} \& {Kroupa}(2009)}]{Pflamm-Altenburg+2009}
{Pflamm-Altenburg}, J., \& {Kroupa}, P. 2009, \apj, 706, 516,
  \dodoi{10.1088/0004-637X/706/1/516}

\bibitem[{{Rey} {et~al.}(2020){Rey}, {Pontzen}, {Agertz}, {Orkney}, {Read}, \&
  {Rosdahl}}]{Rey+2020}
{Rey}, M.~P., {Pontzen}, A., {Agertz}, O., {et~al.} 2020, \mnras, 497, 1508,
  \dodoi{10.1093/mnras/staa1640}

\bibitem[{{Rey} {et~al.}(2022){Rey}, {Pontzen}, {Agertz}, {Orkney}, {Read},
  {Saintonge}, {Kim}, \& {Das}}]{Rey+2022}
---. 2022, \mnras, 511, 5672, \dodoi{10.1093/mnras/stac502}

\bibitem[{{Rhode} {et~al.}(2013){Rhode}, {Salzer}, {Haurberg}, {Van Sistine},
  {Young}, {Haynes}, {Giovanelli}, {Cannon}, {Skillman}, {McQuinn}, \&
  {Adams}}]{Rhode+2013}
{Rhode}, K.~L., {Salzer}, J.~J., {Haurberg}, N.~C., {et~al.} 2013, \aj, 145,
  149, \dodoi{10.1088/0004-6256/145/6/149}

\bibitem[{{Robitaille} {et~al.}(2020){Robitaille}, {Deil}, \&
  {Ginsburg}}]{reproject}
{Robitaille}, T., {Deil}, C., \& {Ginsburg}, A. 2020, {reproject: Python-based
  astronomical image reprojection}.
\newblock \doeprint{2011.023}

\bibitem[{{Sales} {et~al.}(2022){Sales}, {Wetzel}, \& {Fattahi}}]{Sales+2022}
{Sales}, L.~V., {Wetzel}, A., \& {Fattahi}, A. 2022, Nature Astronomy, 6, 897,
  \dodoi{10.1038/s41550-022-01689-w}

\bibitem[{{S{\'a}nchez} \& {Des Collaboration}(2010)}]{DES}
{S{\'a}nchez}, E., \& {Des Collaboration}. 2010, in Journal of Physics
  Conference Series, Vol. 259, Journal of Physics Conference Series, 012080,
  \dodoi{10.1088/1742-6596/259/1/012080}

\bibitem[{{Sand} {et~al.}(2015){Sand}, {Spekkens}, {Crnojevi{\'c}}, {Hargis},
  {Willman}, {Strader}, \& {Grillmair}}]{Sand+2015b}
{Sand}, D.~J., {Spekkens}, K., {Crnojevi{\'c}}, D., {et~al.} 2015, \apjl, 812,
  L13, \dodoi{10.1088/2041-8205/812/1/L13}

\bibitem[{{Sand} {et~al.}(2012){Sand}, {Strader}, {Willman}, {Zaritsky},
  {McLeod}, {Caldwell}, {Seth}, \& {Olszewski}}]{Sand12}
{Sand}, D.~J., {Strader}, J., {Willman}, B., {et~al.} 2012, \apj, 756, 79,
  \dodoi{10.1088/0004-637X/756/1/79}

\bibitem[{{Sand} {et~al.}(2014){Sand}, {Crnojevi{\'c}}, {Strader}, {Toloba},
  {Simon}, {Caldwell}, {Guhathakurta}, {McLeod}, \& {Seth}}]{sand14}
{Sand}, D.~J., {Crnojevi{\'c}}, D., {Strader}, J., {et~al.} 2014, \apjl, 793,
  L7, \dodoi{10.1088/2041-8205/793/1/L7}

\bibitem[{{Sand} {et~al.}(2022){Sand}, {Mutlu-Pakdil}, {Jones}, {Karunakaran},
  {Wang}, {Yang}, {Chiti}, {Bennet}, {Crnojevi{\'c}}, \&
  {Spekkens}}]{Sand+2022}
{Sand}, D.~J., {Mutlu-Pakdil}, B., {Jones}, M.~G., {et~al.} 2022, \apjl, 935,
  L17, \dodoi{10.3847/2041-8213/ac85ee}

\bibitem[{{Sawala} {et~al.}(2016){Sawala}, {Frenk}, {Fattahi}, {Navarro},
  {Bower}, {Crain}, {Dalla Vecchia}, {Furlong}, {Helly}, {Jenkins}, {Oman},
  {Schaller}, {Schaye}, {Theuns}, {Trayford}, \& {White}}]{Sawala+2016}
{Sawala}, T., {Frenk}, C.~S., {Fattahi}, A., {et~al.} 2016, \mnras, 457, 1931,
  \dodoi{10.1093/mnras/stw145}

\bibitem[{{Schlafly} \& {Finkbeiner}(2011)}]{Schlafly2011}
{Schlafly}, E.~F., \& {Finkbeiner}, D.~P. 2011, \apj, 737, 103,
  \dodoi{10.1088/0004-637X/737/2/103}

\bibitem[{{Schlegel} {et~al.}(1998){Schlegel}, {Finkbeiner}, \&
  {Davis}}]{Schlegel1998}
{Schlegel}, D.~J., {Finkbeiner}, D.~P., \& {Davis}, M. 1998, \apj, 500, 525,
  \dodoi{10.1086/305772}

\bibitem[{{Simon}(2019)}]{Simon+2019}
{Simon}, J.~D. 2019, \araa, 57, 375,
  \dodoi{10.1146/annurev-astro-091918-104453}

\bibitem[{{Simpson} {et~al.}(2013){Simpson}, {Bryan}, {Johnston}, {Smith}, {Mac
  Low}, {Sharma}, \& {Tumlinson}}]{Simpson+2013}
{Simpson}, C.~M., {Bryan}, G.~L., {Johnston}, K.~V., {et~al.} 2013, \mnras,
  432, 1989, \dodoi{10.1093/mnras/stt474}

\bibitem[{{Smercina} {et~al.}(2018){Smercina}, {Bell}, {Price}, {D'Souza},
  {Slater}, {Bailin}, {Monachesi}, \& {Nidever}}]{Smercina+2018}
{Smercina}, A., {Bell}, E.~F., {Price}, P.~A., {et~al.} 2018, \apj, 863, 152,
  \dodoi{10.3847/1538-4357/aad2d6}

\bibitem[{{Smercina} {et~al.}(2017){Smercina}, {Bell}, {Slater}, {Price},
  {Bailin}, \& {Monachesi}}]{Smercina+2017}
{Smercina}, A., {Bell}, E.~F., {Slater}, C.~T., {et~al.} 2017, \apjl, 843, L6,
  \dodoi{10.3847/2041-8213/aa78fa}

\bibitem[{{Stetson}(1987)}]{Stetson87}
{Stetson}, P.~B. 1987, \pasp, 99, 191, \dodoi{10.1086/131977}

\bibitem[{{Stetson}(1994)}]{Stetson94}
---. 1994, \pasp, 106, 250, \dodoi{10.1086/133378}

\bibitem[{{Taylor} {et~al.}(2011){Taylor}, {Hopkins}, {Baldry}, {Brown},
  {Driver}, {Kelvin}, {Hill}, {Robotham}, {Bland-Hawthorn}, {Jones}, {Sharp},
  {Thomas}, {Liske}, {Loveday}, {Norberg}, {Peacock}, {Bamford}, {Brough},
  {Colless}, {Cameron}, {Conselice}, {Croom}, {Frenk}, {Gunawardhana},
  {Kuijken}, {Nichol}, {Parkinson}, {Phillipps}, {Pimbblet}, {Popescu},
  {Prescott}, {Sutherland}, {Tuffs}, {van Kampen}, \&
  {Wijesinghe}}]{Taylor+2011}
{Taylor}, E.~N., {Hopkins}, A.~M., {Baldry}, I.~K., {et~al.} 2011, \mnras, 418,
  1587, \dodoi{10.1111/j.1365-2966.2011.19536.x}

\bibitem[{{Tollerud} \& {Peek}(2018)}]{Tollerud+2018}
{Tollerud}, E.~J., \& {Peek}, J.~E.~G. 2018, \apj, 857, 45,
  \dodoi{10.3847/1538-4357/aab3e4}

\bibitem[{{van der Walt} {et~al.}(2011){van der Walt}, {Colbert}, \&
  {Varoquaux}}]{numpy}
{van der Walt}, S., {Colbert}, S.~C., \& {Varoquaux}, G. 2011, Computing in
  Science and Engineering, 13, 22, \dodoi{10.1109/MCSE.2011.37}

\bibitem[{{W}es {M}c{K}inney(2010)}]{pandas1}
{W}es {M}c{K}inney. 2010, in {P}roceedings of the 9th {P}ython in {S}cience
  {C}onference, ed. {S}t\'efan van~der {W}alt \& {J}arrod {M}illman, 56 -- 61,
  \dodoi{10.25080/Majora-92bf1922-00a}

\bibitem[{{Wetzel} {et~al.}(2016){Wetzel}, {Hopkins}, {Kim},
  {Faucher-Gigu{\`e}re}, {Kere{\v{s}}}, \& {Quataert}}]{Wetzel+2016}
{Wetzel}, A.~R., {Hopkins}, P.~F., {Kim}, J.-h., {et~al.} 2016, \apjl, 827,
  L23, \dodoi{10.3847/2041-8205/827/2/L23}

\bibitem[{{Wheeler} {et~al.}(2015){Wheeler}, {O{\~n}orbe}, {Bullock},
  {Boylan-Kolchin}, {Elbert}, {Garrison-Kimmel}, {Hopkins}, \&
  {Kere{\v{s}}}}]{Wheeler+2015}
{Wheeler}, C., {O{\~n}orbe}, J., {Bullock}, J.~S., {et~al.} 2015, \mnras, 453,
  1305, \dodoi{10.1093/mnras/stv1691}

\bibitem[{{Willman} {et~al.}(2005){Willman}, {Dalcanton}, {Martinez-Delgado},
  {West}, {Blanton}, {Hogg}, {Barentine}, {Brewington}, {Harvanek}, {Kleinman},
  {Krzesinski}, {Long}, {Neilsen}, {Nitta}, \& {Snedden}}]{Willman+2005}
{Willman}, B., {Dalcanton}, J.~J., {Martinez-Delgado}, D., {et~al.} 2005,
  \apjl, 626, L85, \dodoi{10.1086/431760}

\bibitem[{{York} {et~al.}(2000){York}, {Adelman}, {Anderson}, {Anderson},
  {Annis}, {Bahcall}, {Bakken}, {Barkhouser}, {Bastian}, {Berman}, {Boroski},
  {Bracker}, {Briegel}, {Briggs}, {Brinkmann}, {Brunner}, {Burles}, {Carey},
  {Carr}, {Castander}, {Chen}, {Colestock}, {Connolly}, {Crocker}, {Csabai},
  {Czarapata}, {Davis}, {Doi}, {Dombeck}, {Eisenstein}, {Ellman}, {Elms},
  {Evans}, {Fan}, {Federwitz}, {Fiscelli}, {Friedman}, {Frieman}, {Fukugita},
  {Gillespie}, {Gunn}, {Gurbani}, {de Haas}, {Haldeman}, {Harris}, {Hayes},
  {Heckman}, {Hennessy}, {Hindsley}, {Holm}, {Holmgren}, {Huang}, {Hull},
  {Husby}, {Ichikawa}, {Ichikawa}, {Ivezi{\'c}}, {Kent}, {Kim}, {Kinney},
  {Klaene}, {Kleinman}, {Kleinman}, {Knapp}, {Korienek}, {Kron}, {Kunszt},
  {Lamb}, {Lee}, {Leger}, {Limmongkol}, {Lindenmeyer}, {Long}, {Loomis},
  {Loveday}, {Lucinio}, {Lupton}, {MacKinnon}, {Mannery}, {Mantsch}, {Margon},
  {McGehee}, {McKay}, {Meiksin}, {Merelli}, {Monet}, {Munn}, {Narayanan},
  {Nash}, {Neilsen}, {Neswold}, {Newberg}, {Nichol}, {Nicinski}, {Nonino},
  {Okada}, {Okamura}, {Ostriker}, {Owen}, {Pauls}, {Peoples}, {Peterson},
  {Petravick}, {Pier}, {Pope}, {Pordes}, {Prosapio}, {Rechenmacher}, {Quinn},
  {Richards}, {Richmond}, {Rivetta}, {Rockosi}, {Ruthmansdorfer}, {Sandford},
  {Schlegel}, {Schneider}, {Sekiguchi}, {Sergey}, {Shimasaku}, {Siegmund},
  {Smee}, {Smith}, {Snedden}, {Stone}, {Stoughton}, {Strauss}, {Stubbs},
  {SubbaRao}, {Szalay}, {Szapudi}, {Szokoly}, {Thakar}, {Tremonti}, {Tucker},
  {Uomoto}, {Vanden Berk}, {Vogeley}, {Waddell}, {Wang}, {Watanabe},
  {Weinberg}, {Yanny}, {Yasuda}, \& {SDSS Collaboration}}]{York+2000}
{York}, D.~G., {Adelman}, J., {Anderson}, John~E., J., {et~al.} 2000, \aj, 120,
  1579, \dodoi{10.1086/301513}

\bibitem[{{Zaritsky} {et~al.}(2023){Zaritsky}, {Donnerstein}, {Dey},
  {Karunakaran}, {Kadowaki}, {Khim}, {Spekkens}, \& {Zhang}}]{Zaritsky+2023}
{Zaritsky}, D., {Donnerstein}, R., {Dey}, A., {et~al.} 2023, \apjs, 267, 27,
  \dodoi{10.3847/1538-4365/acdd71}

\bibitem[{{Zaritsky} {et~al.}(2021){Zaritsky}, {Donnerstein}, {Karunakaran},
  {Barbosa}, {Dey}, {Kadowaki}, {Spekkens}, \& {Zhang}}]{Zaritsky+2021}
{Zaritsky}, D., {Donnerstein}, R., {Karunakaran}, A., {et~al.} 2021, \apjs,
  257, 60, \dodoi{10.3847/1538-4365/ac2607}

\bibitem[{{Zaritsky} {et~al.}(2022){Zaritsky}, {Donnerstein}, {Karunakaran},
  {Barbosa}, {Dey}, {Kadowaki}, {Spekkens}, \& {Zhang}}]{Zaritsky+2022}
---. 2022, \apjs, 261, 11, \dodoi{10.3847/1538-4365/ac6ceb}

\bibitem[{{Zaritsky} {et~al.}(2019){Zaritsky}, {Donnerstein}, {Dey},
  {Kadowaki}, {Zhang}, {Karunakaran}, {Mart{\'\i}nez-Delgado}, {Rahman}, \&
  {Spekkens}}]{Zaritsky+2019}
{Zaritsky}, D., {Donnerstein}, R., {Dey}, A., {et~al.} 2019, \apjs, 240, 1,
  \dodoi{10.3847/1538-4365/aaefe9}

\bibitem[{{Zhang} {et~al.}(2021){Zhang}, {Wu}, {Li}, {Tsai}, {Staveley-Smith},
  {Wang}, {Fu}, {McIntyre}, \& {Yuan}}]{Zhang+2021}
{Zhang}, K., {Wu}, J., {Li}, D., {et~al.} 2021, \mnras, 500, 1741,
  \dodoi{10.1093/mnras/staa3275}

\bibitem[{{Zibetti} {et~al.}(2009){Zibetti}, {Charlot}, \&
  {Rix}}]{Zibetti+2009}
{Zibetti}, S., {Charlot}, S., \& {Rix}, H.-W. 2009, \mnras, 400, 1181,
  \dodoi{10.1111/j.1365-2966.2009.15528.x}

\end{thebibliography}
\bibliographystyle{aasjournal}



\end{document}